\documentclass[pra,twocolumn,showpacs,preprintnumbers,amsmath,amssymb]{revtex4-1}

\usepackage{graphicx}
\usepackage{dcolumn}
\usepackage{bm}
\usepackage{braket}
\usepackage{comment}

\usepackage[normalem]{ulem} 			

\begin{document}

\title{Orbital magnetism of ultracold fermionic gases in a lattice: \\ Dynamical mean-field approach
}

\author{Agnieszka Cichy$^1$}
\author{Andrii Sotnikov$^{2,3}$}
\affiliation{$^{1}$Institut f\"ur Theoretische Physik, Goethe-Universit\"at, 60438 Frankfurt am Main, Germany}
\affiliation{$^{2}$Akhiezer Institute for Theoretical Physics, NSC KIPT, 61108 Kharkiv, Ukraine}
\affiliation{$^{3}$Institute of Physics, ASCR, 18221 Praha 8, Czech Republic}

\date{23 May 2016}

\begin{abstract}
We study finite-temperature properties of ultracold four-component mixtures of alkaline-earth-metal-like atoms in optical lattices that can be effectively described by the two-band spin-$1/2$ Hubbard model including Hund's exchange coupling term. Our main goal is to investigate the effect of exchange interactions on finite-temperature magnetic phases for a wide range of lattice fillings. We use the dynamical mean-field theory approach and its real-space generalization to obtain finite-temperature phase diagrams including transitions to magnetically ordered phases. It allows to determine optimal experimental regimes for approaching long-range ferromagnetic ordering in ultracold gases. We also calculate the entropy 
in the vicinity of magnetically ordered phases, which provides quantitative predictions for ongoing and future experiments aiming at approaching and studying long-range ordered states in optical lattices.

\end{abstract}

\maketitle

\section{Introduction}

Quantum simulators with fully controllable interaction parameters,
which can describe fundamental phenomena in condensed-matter physics, have been
a goal of the solid-state-physics community for over 30 years. Such simulators can provide information on 
the main properties of the physical system in question and isolate different effects and mechanisms, which, in general, constitutes a challenging task to realize in bulk materials due to their inevitable complexity.  

Experiments in which fermionic or bosonic gases are loaded into optical
lattices have already been carried out \cite{greiner, chin, koln, stoferle, Blo2008RMP, Jor2010PRL, SchneiderThesis} and have led to further developments of the field. Both the depth of the
periodic trapping potential and the lattice geometry can be fully controlled. In
this way, a wide range of strongly correlated systems with different geometries can be investigated. Therefore, atomic gases with tunable interactions in optical
lattices allow experimental realization of the Hubbard model in different regimes with ultimate control and tunability.

The simple electronic structure of alkali-metal atoms (a single valence electron) in combination with Feshbach-resonances sets new
perspectives for experimental realization and study of many different unconventional
systems, including superconductivity with nontrivial Cooper pairing \cite{fulde, Larkin, Bianchi, Nardulli, Matsuda, torma, koponen, Mierzejewski, loh},  Bose-Fermi mixtures \cite{MicnasModern, ketterle, ketterle2, titvinidze, titvinidze2}, and mixtures of fermions with three hyperfine states (in analogy to quantum chromodynamics) \cite{Hofstetter, privitera}. However, relative simplicity of the internal structure of these atoms also introduces many limitations and implies that some phenomena can not be directly observed within current experiments working with ultacold gases of alkali-metal atoms \cite{Rey_review}. So far, only short-range magnetic correlations have been observed \cite{Greif, Har2015Nat} in the Fermi and Bose-Hubbard models because the energy scales of the effective long-range spin-spin correlations set low entropy requirements.

On the other hand, remarkable developments of experimental techniques in ultracold gases of alkaline-earth-metal-like atoms \cite{Takahashi, Taie, Thobe, Cappellini, Pagano, Scazza} in recent years have allowed investigation of new states of matter that go beyond the possibilities already offered by conventional condensed-matter systems \cite{Gorshkov, Feig}. The unique properties of these atoms \cite{Thobe} also give an opportunity to study Hamiltonians beyond the reach of current alkali-based experiments. Among these are the following: the two-band Hubbard model revealing, in particular, the physics of the orbital-selective Mott transition \cite{Jakobi}, the Kugel-Khomskii model originally proposed for transition-metal oxides \cite{Kugel}, the Kondo lattice model studied in the context of manganese oxide perovskites and heavy fermion materials
\cite{Rey2009}, and SU($N$)-symmetric magnetic systems \cite{Hofstetter, Sotnikov}. Ytterbium is particularly versatile due to its large number of bosonic and fermionic isotopes with a wide range of interaction strengths \cite{Scazza} and also has some advantages, such as low-lying (meta)stable excited electronic states, decoupling of the nuclear spin from the electronic
degrees of freedom, and different ac-polarizabilities of the ground and metastable states.

A central topic in quantum magnetism is the interplay between orbital and spin degrees of freedom. As pointed out in Ref.~\cite{Gorshkov}, systems consisting of alkaline-earth-metal-like atoms are most convenient for realizations and studies of the two-band Hubbard model over a wide parameter range.
One interesting aspect of these systems is the possibility of magnetic ordering of nuclear spins. Therefore, one of the main goals of our paper is to determine parameter regimes that are most favorable for ferromagnetic (FM) phases, so that they can be observed within ultracold atomic gases in optical lattices at experimentally accessible values of temperatures and entropies. Our expectations in this respect are based on the double-exchange mechanism \cite{Zener}. Its physics is most pronounced in the Kondo lattice model, which is also often referred to as the double-exchange model.

This paper is organized as follows. In Sec.~\ref{sec.2}, we introduce the 
main theoretical model for the system under study, the two-orbital Hubbard model, and briefly discuss some relevant limiting cases. In Sec.~\ref{sec.3}, we briefly present our numerical method, the dynamical mean-field approach. 
Sec.~\ref{sec.4} presents the numerical results and a discussion of the magnetic properties (Sec.~\ref{subsec.4a}) and the entropy analysis (Sec.~\ref{subsec.4b}). 
We conclude in Sec.~\ref{sec.5} with a brief summary of the obtained results and an outlook.

\section{System and Model}\label{sec.2}
We consider a two-orbital Fermi-Hubbard Hamiltonian of the following type \cite{Gorshkov}:
\begin{eqnarray}\label{Hubb_two_orb}
H&=&-t\sum_{\langle i,j \rangle \alpha \sigma} (c_{i\alpha \sigma}^{\dag} c_{j\alpha \sigma}+H.c.)+U\sum_{i,\alpha} n_{i\alpha \uparrow}n_{i\alpha \downarrow}\nonumber \\
&+& \sum_{i \sigma \sigma'}(V-\delta_{\sigma \sigma'}V_{ex}) n_{ie\sigma}n_{ig\sigma'}\nonumber \\
&+& V_{ex}\sum_{i,\sigma\neq \sigma'} c_{ig\sigma}^{\dag} c_{ie\sigma'}^{\dag} c_{ig\sigma'}c_{ie\sigma}-\sum_{i\alpha}\mu_{\alpha}n_{i\alpha},
\end{eqnarray}
where $\alpha=\{g,e\}$ denotes the electronic state representing the orbital, $\sigma=\{-I,\dots, I\}$ denotes one of the $N=2I+1$ nuclear Zeeman states representing the spin, the operator $c_{i\alpha \sigma}^{\dag}$ ($c_{i\alpha \sigma}$) creates (annihilates) an atom in the internal state $|\alpha \sigma \rangle$ at the site $i$, $t$ is the hopping amplitude, $n_{i\alpha \sigma}=c_{i\alpha \sigma}^{\dag} c_{i\alpha \sigma}$, and $n_{i\alpha}=\sum_{\sigma} n_{i\alpha \sigma}$. 
The notation $\langle i,j \rangle$ indicates the summation over nearest-neighbor sites. $U=g_{\alpha \alpha} \int d^3 r w_{\alpha}^4 ({\bf r})$ is the Hubbard interaction within a single band (only $g$ or $e$ atoms), and $V=(U_{eg}^{+}+U_{eg}^{-})/2$ and $V_{ex}=(U_{eg}^{+}-U_{eg}^{-})/2$ describe the on-site direct and exchange interaction terms, respectively, with $g_{\alpha\alpha'}=4\pi a_{\alpha\alpha'}/M$, where $a_{\alpha\alpha'}$ is the scattering length of two atoms on orbitals $\alpha$ and $\alpha'$, $M$ is the atomic mass, $w_\alpha({\bf r})$ is the Wannier function on
orbital $\alpha$, and $U_{eg}^{\pm}=g_{eg}^{\pm} \int d^3 r w_{e}^{2}({\bf r}) w_{g}^{2}({\bf r})$. 
Note that the inter-orbital exchange interaction $V_{ex}$ is separated into its density-density and spin-flip 
contributions [the terms in the second and the third lines of Eq.~(\ref{Hubb_two_orb}), respectively]; the latter is neglected below in our numerical analysis (see Sec.~\ref{sec.3}).   

The system described by the Hamiltonian~(\ref{Hubb_two_orb}) can be experimentally realized by loading ground-state atoms (denoted by $|g\rangle$) that are prepared in two different nuclear Zeeman states in the optical lattice, i.e., the realization that is directly related to the solid-state materials, $\sigma, \sigma'=\{\uparrow, \downarrow\}$. 
The system can be extended to the second orbital by the supplementary electronic state $|e\rangle$. 
Appropriate candidates are mixtures of alkaline-earth-metal-like atoms (in particular, $^{173}$Yb and $^{87}$Sr fermionic isotopes, see, e.g., Refs.~\cite{Takahashi,Ste2011PRA,Scazza,Cappellini}) due to the presence of two stable electronic states, $^{1}S_0=|g\rangle$ and $^{3}P_0=|e\rangle$. The different ac polarizability
of the ground- and (meta)stable states also gives an opportunity to realize the Kondo lattice model in ultracold gases. 

Let us note that for the amplitude of the lattice potential $V_0\approx3E_{r}$, with $E_r=\hbar^2 k^2/(2m)$ being the recoil energy of the atoms, typical ratios of the Hubbard parameters are $V_{ex}/U\sim V/U\sim 10$ under the conditions of the experiment~\cite{Scazza,Note1}. These ratios decrease with an increase in the lattice potential $V_0$. Hence, we focus on the case of a sufficiently strong lattice potential $V_0\gtrsim5E_r$ in our analysis. Due to experimental limitations (e.g., large particle losses related to the three-body collisions in weak lattice potentials), we choose $U=12t$, $V=22t$, and $V_{ex}=21t$ as the central values~\footnote{%
    Private communication with S. F{\"o}lling, F. Scazza, D. R. Fernandes and L. Riegger, LMU, Munich.}. 
    However, according to the mentioned tunability by applying the Feshbach-resonance technique in ultracold atoms, in Sec.~\ref{sec.4} we show results assuming that one or two parameters can be additionally tuned with respect to these central values. Therefore, we also provide important predictions for the most optimal regimes in approaching and studying the long-range ordered phases in optical lattices.

An important insight into many-body states and possible magnetic phases of the model~(\ref{Hubb_two_orb}) can be gained by considering the strong-coupling (SC) limit. For definiteness, assuming all the constants entering Eq.~(\ref{Hubb_two_orb}) are positive, $\Delta=(\mu_g-\mu_e)\geq0$ (this corresponds to the condition for the total number of particles $N_g\geq N_e$ in experiments), and $\Gamma=\min\{U, \,V, \,V_{ex}\}$, the SC limit can be associated with the condition $\Gamma \gg t$. Note that, in contrast to the single-band Fermi-Hubbard Hamiltonian, the model~(\ref{Hubb_two_orb}) is more complex due to the wider range of lattice fillings and parameters involved. Nevertheless, the corresponding SC limits at different fillings have been studied in the literature in different contexts (see, e.g., Refs.~\cite{Jakobi,Held,Peters,Kunes} and references therein). In particular, at quarter filling ($n= \langle n_{ie} + n_{ig}\rangle\approx1$), the system can be mapped to the effective spin-1/2 Heisenberg model: 
(a) 
with 
the antiferromagnetic coupling between two spin states $\ket{\uparrow}$ and $\ket{\downarrow}$ within the $g$ orbital at $\Gamma<\Delta$, or (b) with the antiferroorbital coupling between the same $\sigma$ states on different orbitals at $\Delta\lesssim t\ll \Gamma$. In turn, at half filling ($n\approx2$), the system can be mapped to the effective spin-1 Heisenberg model with positive (``antiferromagnetic'') effective coupling constants between adjacent ``spins'' (hard-core bosons) that can be expressed in terms of creation and annihilation operators of fermionic double occupancies. Finally, at non-integer values of lattice fillings (at $n<1$ and $1<n<2$) the leading contributions are expected to come from the double-exchange mechanism \cite{Zener}; thus, the ferromagnetic order can develop in these regions depending on the explicit values of $n$, $\Delta$, $U$, $V$, and $V_{ex}$.

\section{Method}\label{sec.3}
Our theoretical analysis is based on the dynamical mean-field theory (DMFT) numerical approach \cite{Georges, Vollhardt}. Although a number of relevant properties of strongly correlated systems can be studied by using the standard mean-field approximation, this method is certainly limited. For instance, the Hartree-Fock approximation usually significantly overestimates the critical temperatures for the long-range ordered phases in the intermediate- and strong-coupling limits (see, e.g., Ref.~\cite{Held}). In contrast, the DMFT technique allows us to retain
all the dynamical quantum fluctuations of the self-consistent local effective
problem onto which, as in standard mean-field theories, the full lattice problem
is effectively mapped \cite{Isidori}. Since all the energy scales characterizing the effective problem
are treated on the same level, this scheme is considered to be one of the most powerful theoretical approaches to the
non-perturbative regime of the Hubbard model and other strongly correlated
systems.

The physical idea is based on an approximation of the full lattice problem with a single-site effective problem. The effects of the rest of the lattice on each site are described by a self-consistent effective field (analogous to the static Weiss field) acting on the local degrees of freedom of the single-site problem. The local dynamics of the DMFT effective problem is treated exactly (now, the Weiss field is transformed to its dynamical version).
The DMFT approximation becomes exact in the limit of infinite dimensions (i.e., the limit of infinite lattice coordination number $z$). In this limit, the momentum conservation is irrelevant in the interaction vertices of all self-energy diagrams. Hence, the total self-energy $\Sigma (\omega)$ is momentum independent. The locality of the self-energy becomes the main starting point in the development of DMFT. An auxiliary impurity problem [usually, the Anderson impurity model (AIM) \cite{Anderson}] and its action $S_{eff}$ are introduced, containing an impurity with the same local interactions as on the lattice (the impurity is connected to a non-interacting bath).
Although it is not an exact method in the case of square and cubic lattice geometries ($z=4$ and $z=6$, respectively), results obtained with DMFT are often considered as a good reference point both for experiments and for other numerical approaches. 

The most challenging part of the DMFT procedure is to solve the impurity problem. For the case under study, the AIM Hamiltonian can be written in the following form:
\begin{eqnarray}
\label{Ham_AIM}
H_{AIM}&=&\sum_{\alpha \sigma} \sum_{l=2}^{n_s} 
\Bigl[
\epsilon_{l\alpha \sigma}a_{l\alpha \sigma}^{\dag} a_{l\alpha \sigma} +  V_{l\alpha \sigma} (d_{\alpha \sigma}^{\dag} a_{l\alpha \sigma}+H.c.)
\Bigr]
\nonumber \\
&+&U  \sum_{\alpha} n_{\alpha \uparrow}^{d}n_{\alpha \downarrow}^{d}
+ \sum_{\sigma \sigma'}(V-\delta_{\sigma \sigma'}V_{ex}) n_{e\sigma}^{d}n_{g\sigma'}^{d}\nonumber \\
&-&\sum_{\alpha}\mu_{\alpha}n_{\alpha}^{d},
\end{eqnarray}
where the spin-flip term from Eq.~(\ref{Hubb_two_orb}) is omitted, $d_{\alpha \sigma}^{\dag}$ and $a_{\alpha \sigma}^{\dag}$ create the ``impurity'' and bath electrons, respectively, and the index $l$ labels the number of the bath's orbital, with $n_s$ being the cut-off number peculiar to the exact diagonalization (ED) approach \cite{Georges} (one of the methods to solve the AIM problem). The quantities $\epsilon_{l\alpha \sigma}$ and $V_{l\alpha \sigma}$ (bath's energies and hybridization amplitudes, respectively) are the so-called Anderson parameters that are determined self-consistently within the DMFT iterative procedure.
To solve the effective quantum impurity problem, we use the ED method \cite{Georges, Caffarel, Sotnikov2} as well as the continuous-time quantum Monte-Carlo hybridization expansion solver (CT-HYB) (see Refs. \cite{Sotnikov, Gull, Buchhold} for details).

The impurity problem is related to the lattice problem through the self-consistency
condition (the Dyson equation) 
\begin{eqnarray}\label{Dyson}
\Sigma_{\alpha \sigma}(i\omega_n) = {\cal G}^{-1}_{\alpha \sigma}(i\omega_n) - {G}^{-1}_{\alpha \sigma}(i\omega_n),
\end{eqnarray}
where ${\cal G}_{\alpha \sigma}(i\omega_n)$ is the Weiss Green's function, ${G}_{\alpha \sigma}(i\omega_n)$ is the local lattice Green's function, $\omega_n=(2n+1)\pi/\beta$ is the fermionic Matsubara frequency, $n$ is an integer number, and $\beta$ is the inverse temperature in units of $k_B=1$. In turn, the Weiss Green's function can be directly expressed in terms of the hybridization function~$\Delta_{\alpha \sigma}(i\omega_n)$ as ${\cal G}^{-1}_{\alpha \sigma}(i\omega_n)=i\omega_n+\mu_\alpha -\Delta_{\alpha \sigma}(i\omega_n)$.
Therefore, condition~(\ref{Dyson}) relates the hybridization function $\Delta_{\alpha \sigma}(i\omega_n)$ that is determined by the Anderson parameters as 
\begin{equation}
\Delta_{\alpha \sigma}(i\omega_n)=\sum_{l=2}^{n_s}\frac{|V_{l\alpha \sigma}|^{2}}{i\omega_n-\epsilon_{l\alpha \sigma}} 
\end{equation}
and the local Green's function at the given frequency~$\omega_n$. 

\subsection{Single-site DMFT and FM ordering analysis}

For the multi-band Hubbard model as given by Eq.~(\ref{Hubb_two_orb}), the DMFT self-consistency condition for homogeneous phases in the presence of the magnetic field~$h$ (necessary below for the analysis of the FM-ordered phases) can be written as follows:
\begin{equation}
\label{ferro}
G_{\alpha\sigma}(i\omega_n)=\int d\epsilon \frac{D(\epsilon)}{i\omega_n+\mu_{\alpha}+h\sigma-\epsilon -\Sigma_{\alpha\sigma}(i\omega_n)}, 
\end{equation}
where $D(\epsilon)$ is the non-interacting density of states (we consider the simple cubic lattice geometry below).
The magnetization as a function of the external field~$h$ is
given by
\begin{equation}
M(h)=\frac{1}{\beta}\sum_{\alpha} \sum_n e^{i\omega_n 0^+} \left[G_{\alpha \uparrow}(i\omega_n)-G_{\alpha\downarrow}(i\omega_n)\right],
\end{equation}
where the field $h$ enters implicitly via the Green's functions dependence on $h$.

A ferromagnetic phase manifests itself by a non-zero spontaneous magnetization $M_0=\lim_{h\rightarrow 0}M(h)\neq 0 $. 
The fits for the magnetic susceptibility $\chi$ read
\begin{equation}\label{resid}
M(h)=M_0+\chi h.  
\end{equation}
Since the magnetic susceptibility diverges at the critical temperature~$T_{\text{C}}$, the phase boundaries can be determined by the Curie--Weiss fit
\begin{equation}\label{CWfit}
\chi=\frac{c}{T-T_{\text{C}}}, 
\end{equation}
where $c$ is a fitting parameter.

\subsection{Two-sublattice and real-space DMFT}

To allow for antiferromagnetic (AFM) ordering of the bipartite type, the DMFT self-consistency conditions must be extended to the following form:
\begin{equation}
G_{\alpha\sigma s}(i\omega_n)=\int \frac{\zeta_{\alpha\sigma \bar{s}}D(\epsilon)d\epsilon}{\zeta_{\alpha\sigma A}\zeta_{\alpha\sigma B}-\epsilon^2}  
\end{equation}
with $\zeta_{\alpha\sigma s}\equiv i\omega_n +\mu_\alpha-\Sigma_{\alpha\sigma s}(i\omega_n)$ and the sublattice indices $s=A, B$ and their opposites $\bar{s}=B, A$.

However, the single-site or two-sublattice DMFT self-consistency conditions introduced above are obviously not enough to capture more exotic types of magnetic order (see, e.g., Ref.~\cite{Sotnikov2}). In this case, the real-space generalization of DMFT (RDMFT) is required \cite{Helmes, Snoek}. The RDMFT approximation is successfully used for the system that has an enlarged unit cell, i.e., spanning more than two lattice sites, or even without any translational symmetry remaining. Within this approach, the local self-energy depends on the spatial index that substantially enlarges the required computational resources. However, this limitation can be significantly reduced for systems with periodic order in the case when the explicit type of ordering is known beforehand (e.g., from the unbiased 
RDMFT analysis). The \emph{so-called} clustering procedure in RDMFT is discussed in detail in Ref. \cite{Sotnikov2} in the context of three-sublattice AFM ordering in the SU(3)-symmetric Hubbard model at filling $n=1$ and can be applied to the system under study in a straightforward way.  

In the present paper, numerical results are obtained by using two types of impurity solvers: ED (limited to $n_s=4$ per each $\ket{\alpha\sigma}$ state) and CT-HYB. In addition to the single-site and two-sublattice versions, we also applied the RDMFT approach (typically for system size not larger than $12^3$ sites) where it was necessary. 

\section{Results}\label{sec.4}
\subsection{Magnetic properties}\label{subsec.4a}

In this section, we focus on the analysis of magnetic properties
of ultracold atomic mixtures assuming the simple cubic lattice geometry (as in the experiments~\cite{Pagano, Scazza}). Within the DMFT
approach (and its case-specific modifications discussed in Sec.~\ref{sec.3}), we construct phase diagrams in two ways: by fixing the chemical
potential~$\mu$ (the natural input variable in the numerical analysis) or the lattice filling~$n$ (density of atoms per site, which is more relevant for experiments). Since the Hund's coupling amplitude $V_{ex}$, as well as interaction strengths $V$ and $U$, can be tuned with a high level of control in experiments working with ultracold gases, our analysis is focused on finding the optimal parameter regime for approaching magnetic phases. Note that we use slightly different parameter ranges than the ones that are typically used for modeling solid-state systems \cite{Held}. 
The reason is that our choice for the main parameters is based on the experiments~\cite{Pagano, Scazza}, where the spin-changing contact interactions were realized.
There, an exchange between particles in the two separate orbitals is mediated by the contact interaction between atoms that is characterized by the clock-shift spectroscopy in a three-dimensional optical lattice \cite{Scazza}.
All relevant scattering channels
for atom pairs in combinations of the ground and the excited states can also be effectively determined by the corresponding measurements. In particular, an elastic scattering between
the orbitals is dominated by the antisymmetric channel that results in the extremely strong spin-exchange coupling. 
Our ranges for the values of $U$, $V$, and $V_{ex}$ are chosen to correspond to this experimental situation (see Sec.~\ref{sec.2} for more details). 

First, we focus on the analysis of the finite-temperature phase diagrams for fixed chemical potential $\mu$ in the region of $n\approx1.5$.
To solve the effective quantum impurity problem within DMFT, we choose the CT-HYB solver (which is found to be reliable down to $T\approx0.05$) for this case. Note that here and below we use the units of $t=1$ for all energy-related quantities.
Below the critical temperature (the Curie temperature $T_{\text{C}}$), due to spontaneous symmetry breaking, the magnetic moments on each lattice site become aligned with their neighbors. The magnetic susceptibility $\chi$ diverges at $T_{\text{C}}$ in the thermodynamic limit. Hence, the most direct method is to analyze $\chi$ and thus determine critical temperature for the FM ordering. 

\begin{figure}
\includegraphics[width=\linewidth]{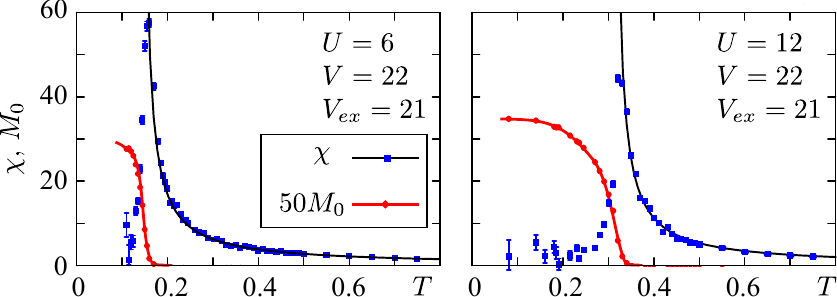}
    \caption{\label{fig1} (color online)
    Magnetic susceptibility $\chi$ and residual magnetization $M_0$ as functions of temperature $T$ for fixed $\mu=2.5$, $V=22$, $V_{ex}=21$ and two different values $U$, $U=6$ (left) and $U=12$ (right).
    }
\end{figure}
In Fig. \ref{fig1} we show the magnetic susceptibility~$\chi$ and the residual magnetization~$M_0$ as functions of the temperature~$T$ at fixed $\mu=2.5$, $V=22$, $V_{ex}=21$, and two different values of intraband interactions, $U=6$ and $U=12$. The error bars correspond to statistical errors in the numerical computation. The magnetic susceptibility diverges at $T=T_{\text{C}}$, and the FM phase boundaries are determined according to Eq.~(\ref{CWfit}) by the Curie-Weiss fit (solid black line). 
According to Eq.~(\ref{resid}), we also extract the residual magnetization that becomes non-zero in the vicinity of the second-order phase transition from the paramagnetic (PM) to the FM phase and increases with decreasing temperature.

\begin{figure}
\includegraphics[width=\linewidth]{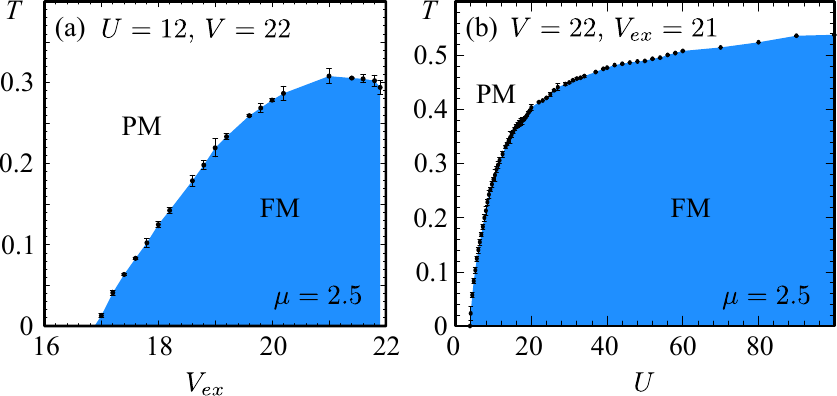}
\caption{\label{fig2} (color online)
Curie temperature as a function of (a) exchange interaction (Ising-type Hund's coupling) at $U=12$ and $V=22$ and (b) intra-orbital interactions at $V=22$ and $V_{ex}=21$. A fixed chemical potential $\mu=2.5$ is chosen to yield $n\approx1.5$ in the PM phase.}
\end{figure}
According to the above scheme, in Fig.~\ref{fig2} we construct the $T-V_{ex}$ and $T-U$ phase diagrams at fixed $\mu=2.5$. A similar analysis was done in Ref.~\cite{Held}; however, there the authors studied the importance of Hund's rule coupling in the presence of orbital degeneracy as a microscopic mechanism of ferromagnetism in transition metals, and the corresponding analysis was limited to the Bethe lattice geometry with symmetric semi-elliptic density of states~$D(\epsilon)$. They found that for intermediate intra- and inter-orbital couplings (i.e., at $U=9$ and $V=5$, respectively) and $n=1.25$, already a small Hund's rule coupling is sufficient to stabilize ferromagnetism.
However, note that for the case of cubic lattice geometry (which is more relevant for optical-lattice experiments) all interaction parameters must be rescaled properly (in particular, due to the three-fold increase of the bandwidth in comparison with \cite{Held}) to allow for the FM order at finite temperature for a wide range of lattice fillings.
While we additionally verified that our DMFT results with the mentioned rescaling qualitatively agree with \cite{Held},
obviously, a more case-specific analysis is required to give reliable quantitative predictions that properly account for the lattice geometry and interaction amplitudes peculiar to the ultracold-atom experiments~\cite{Pagano, Scazza}.

According to Fig.~\ref{fig2}(a), we observe that ferromagnetism is stabilized by the exchange coupling (the Ising-type Hund's coupling $V_{ex}$). 
Note that we also determine a critical value of $V_{ex}$, where the magnetically-ordered phase becomes completely suppressed at $T>0$. With increasing exchange $V_{ex}$, the critical temperature $T_{\text{C}}$ increases and reaches its highest value at $V_{ex}\approx 21$ and then decreases again. It is important to note that the exchange interaction must be smaller than the density-density interaction $V$, since for $V_{ex}>V$ the effective inter-band interaction becomes attractive. Although this region is not considered in the framework of our studies, it remains an interesting research direction from the point of view of the stability of
triplet superconductivity \cite{Kubo}.  

In Fig.~\ref{fig2}(b) we also study the influence of the intra-orbital Hubbard interaction on the stability of the FM phase. With increasing $U$ (at fixed $V=22$ and $V_{ex}=21$), the critical temperature for FM order increases. However, we observe a saturation effect for $T_{\text{C}}$ at large $U$ which directly indicates that FM ordering is governed by double-exchange processes (with an effective amplitude proportional to $V_{ex}=\text{const}$), but not by super-exchange processes (which scale inverse proportionally with the interaction strength~$U$).

\begin{figure}
\includegraphics[width=\linewidth]{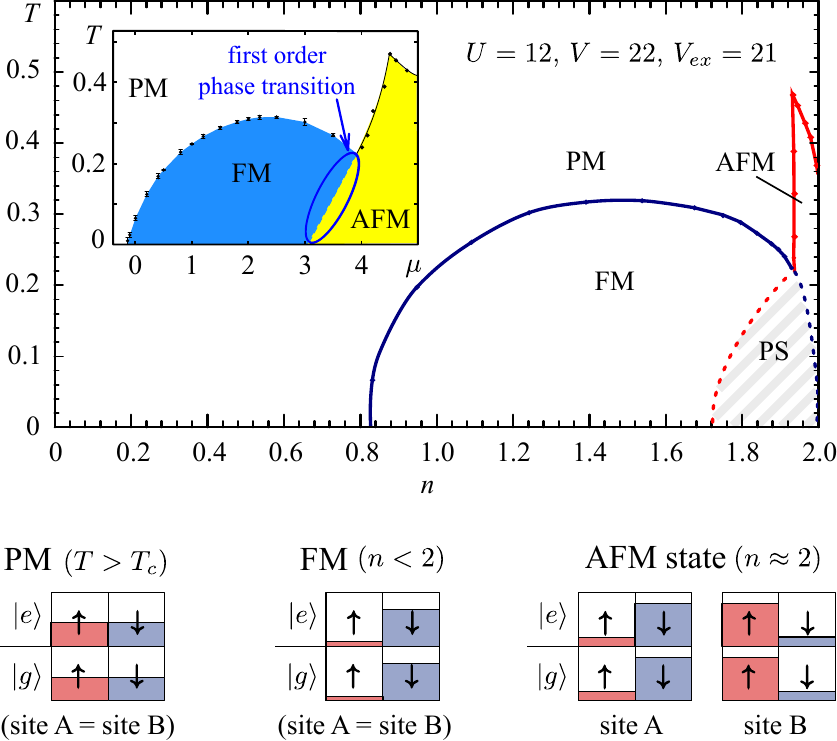}
    \caption{\label{fig5} (color online)
    $T-n$ phase diagram at $U=12$, $V=22$, and $V_{ex}=21$ including magnetically ordered states that are schematically illustrated in the lower part: paramagnetic (PM), ferromagnetic (FM), and antiferromagnetic (AFM); PS denotes the region of phase separation. Inset: $T-\mu$ phase diagram.}
\end{figure}

The above analysis is performed at fixed chemical potential. However, from the point of view of experimental realization of the system with ultracold gases in optical lattices, 
the density of atoms rather than the chemical potential can be directly controlled and tuned.
Hence, it is important to study how the FM phase is affected by a change in the lattice filling~$n$.  
In Fig.~\ref{fig5} we show an example of the $T-n$ phase diagram at $U=12$, $V=22$, and $V_{ex}=21$. 
These calculations were performed by using the ED solver, but the results were also compared with the output from the CT-HYB solver,
so that the critical temperatures for the FM phase are in a good agreement with each other within the statistical error.
From Fig.~\ref{fig5} we conclude that the FM phase is stable for a wide range of lattice fillings, i.e., from $n\approx 0.8$ up to $n\approx 1.9$. 
Due to particle-hole symmetry, 
we only show the range of $n$ from 0 to 2 (half filling). At the chosen interaction parameters, the maximum value of $T_{\text{C}}$ corresponds to $n\approx 1.5$, and typical Curie temperatures are of the order of $T\sim 0.25$. Around half filling, 
ferromagnetic order is suppressed by the antiferromagnetic Heisenberg exchange and two-sublattice AFM order appears in the phase diagram. The corresponding N\'eel temperatures $T_{\text{N}}$ are higher than the Curie temperatures $T_{\text{C}}$. It is important to note that the maximum of $T_{\text{N}}$ is observed at $n\approx 1.94$; that is, it is shifted away from 
the ``undoped'' Mott-insulating state ($n=2$), where the AFM order becomes significantly suppressed at high temperatures. According to additional analysis, we conclude that this behavior is peculiar to the chosen regime $V>V_{ex}>U$. With a decrease of both $V$ and $V_{ex}$ to the values close to or less than the intra-orbital interaction $U$, the more ``conventional'' behavior of $T_{\text{N}}$ is observed with a sharp maximum at $n=2$ that corresponds to a typical scenario in the solid-state realizations \cite{Held}. 
Hence, we observe in this case not only quantitative but also qualitative differences between the solid-state and ultracold gases systems. A schematic illustration of the observed phases in the system under study is also shown in Fig. \ref{fig5}.

We find a phase-separated region (PS) between AFM and FM ordering at low temperatures and $n\gtrsim1.7$ in the $T-n$ phase diagram, which corresponds to the region of the first-order phase transition line in the $T-\mu$ phase diagram (ellipse in the inset of Fig. \ref{fig5}). 
Note that in the numerical analysis of the $T-\mu$ phase diagram in this region, we observe either no stable convergence of DMFT to one or both of the proposed solutions that are peculiar to low temperatures $T\lesssim0.08$ or a reliable convergence to both solutions; thus, a coexistence of two phases is observed at $T\gtrsim0.08$. However, we should emphasize here that the first-order transition indicated in the $T-\mu$ diagram can not be observed directly due to discontinuities in the total number of particles $n$ across the phase boundary at the fixed $\mu$; thus, the only possible mechanism for the system to equilibrate in this region under experimental conditions (fixed $n$) is a phase separation that is observed by our RDMFT analysis.

\begin{figure}
\includegraphics[width=\linewidth]{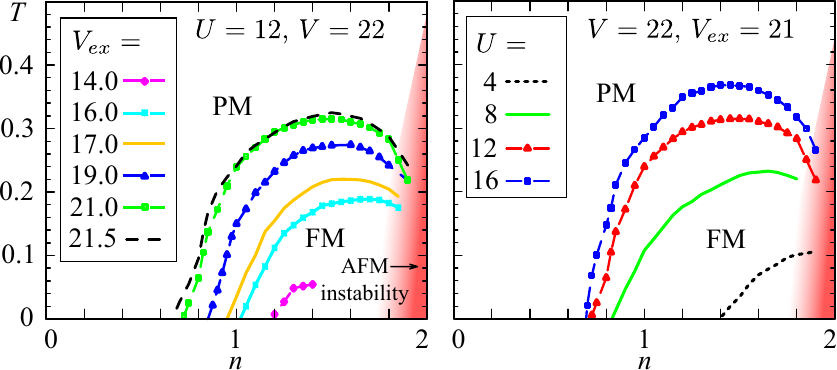}
    \caption{\label{fig6} (color online)
    $T-n$ phase diagram at $U=12$, $V=22$, and different values of the exchange interaction (left) and $V=22$, $V_{ex}=21$, and different values of intra-orbital interactions (right). The AFM instability is indicated here only schematically around half filling.}
\end{figure}

\begin{figure}
\includegraphics[width=\linewidth]{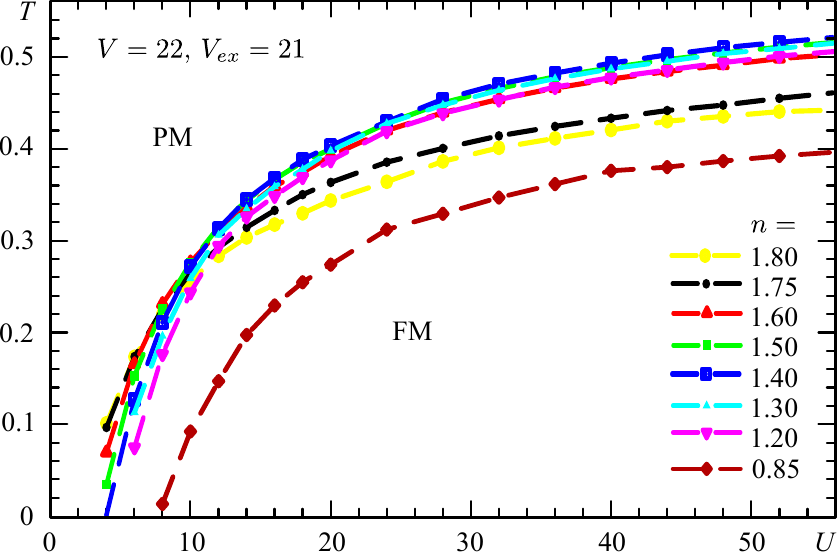}
    \caption{\label{fig8} (color online)
    $T-U$ phase diagram at $V=22$, $V_{ex}=21$ and different values of lattice filling~$n$ obtained by DMFT.}
\end{figure}

Now, let us discuss the influence of the Ising-type Hund's coupling on the stability of the FM phase in more detail. Figure \ref{fig6} shows the  
boundaries of this phase at fixed $U=12$, $V=22$ and six different values of the exchange interaction obtained with the ED solver.  At the highest value of $V_{ex}=21.5$, ferromagnetism is found to be stable for a broad range of particle densities. Note that there exists a critical value of $n$ below which the FM state becomes unstable, in particular, $n_c\approx 0.7$ at $T=0$ and $V_{ex}=21.5$.
This critical value is systematically shifted towards higher $n$ with decreasing exchange interaction. The Curie temperature decreases with decreasing $V_{ex}$ and is significantly lower for $V_{ex}=14$ (the maximum of $T_{\text{C}}\approx 0.06$) than for $V_{ex}=21.5$ (the maximum of $T_{\text{C}}\approx 0.35$). Moreover, the suppression of FM order with decreasing $V_{ex}$ remains strong at higher values of $n$. In this region, the competing AFM instability appears (for simplicity, it is indicated only schematically in Fig.~\ref{fig6} around half filling). 

A further interesting aspect of this analysis is the influence of the intra-orbital interaction on the stability of FM. In Fig.~\ref{fig6} we also show the $T-n$ phase diagram at $V=22$, $V_{ex}=21$ and different values of intra-orbital interaction $U$. It is clearly visible that higher values of $U$ support the ferromagnetic order because of the tendency of the fermions to localize within a single orbital. In the strong-coupling limit, the interactions between particles coming from different orbitals become effectively the interactions between localized moments. 

Finally, in Fig. \ref{fig8} we show the $T-U$ phase diagram at fixed $V=22$, $V_{ex}=21$ and different values of the filling. The computations were done with the ED solver, and the Curie temperatures obtained for $n=1.5$ are in a good agreement with those obtained for $\mu=2.5$ ($n\approx 1.5$) using the CT-HYB solver (see also Fig. \ref{fig2}). In the weak-coupling limit, the critical temperature for FM order increases with $n$. This dependence on $n$ is related to the fact that the Brillouin zone becomes more filled with increasing $n$, and thus, mechanisms for magnetic instabilities are better supported. However, note that at moderate coupling (around $U=10$), the tendency is inverted; that is, the Curie temperature decreases with $n$. In the strong-coupling limit, this behavior can be understood from the argument that the FM state favors some moderate ``doping'' since at larger $n$ it starts to compete with the AFM instability. Therefore, from the perspective of higher $T_{\text{C}}$, the optimal filling 
is 
around $n=1.4$ in the strong-coupling regime ($U>12$).

\subsection{Entropy analysis}\label{subsec.4b}

An important thermodynamical quantity in experiments working with ultracold atomic mixtures in optical lattices is the entropy rather than the temperature. The system is isolated from the environment and experiments are assumed to be performed adiabatically. Hence, the entropy can be considered to be approximately fixed. To calculate the entropy, we use the Maxwell relation for the entropy per site (according to Ref. \cite{Werner}):
\begin{equation}
\label{entropy}
s(\mu_0, U, T)=\int_{-\infty}^{\mu_0} \Bigg(\frac{\partial n (\mu, U, T)}{\partial T} \Bigg) d \mu, 
\end{equation}
where the filling $n$ is the local observable directly measured by DMFT.

\begin{figure}[h!]
\includegraphics[width=\linewidth]{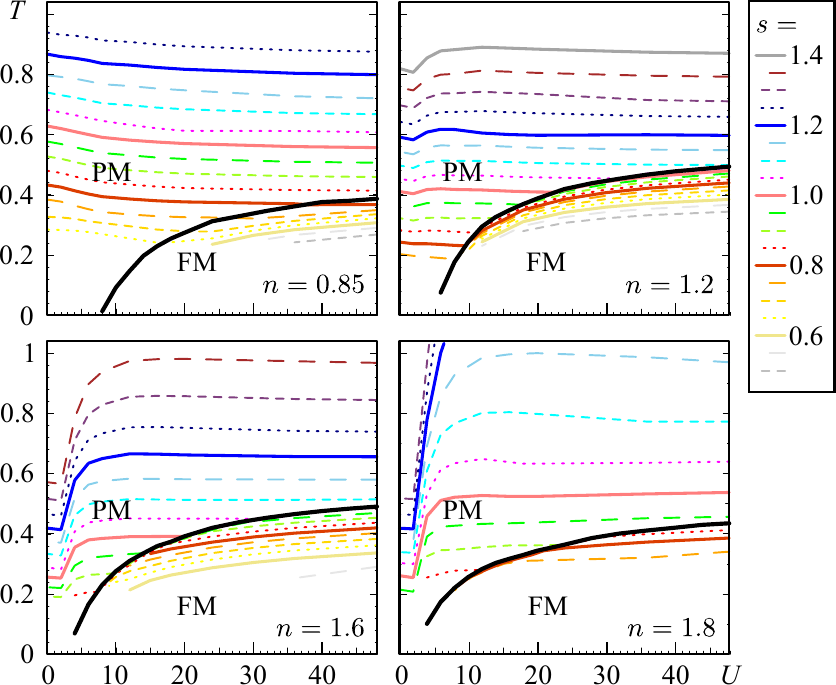}
    \caption{\label{fig9} (color online)
    Isentropic lines at different values of filling: $n=0.85$, $n=1.2$, $n=1.6$, and $n=1.8$. The inter-band interactions are kept fixed, $V=22$ and $V_{ex}=21$.}
\end{figure}

To perform the above integration, the chemical potential is discretized with the parametrization $\mu(r)=\mu_0-V_{tr}r^2$, where $V_{tr}$ is a trapping potential, $r$ is the distance from the trap center and $\mu_0$ is
the chemical potential in the center of the trap \cite{Sotnikov, Sotnikov2, Golubeva}. This parametrization is equivalent to the local-density approximation (LDA) for a system with an external trapping potential. 
Note that, according to our experience, in the numerical integration procedure (\ref{entropy}) a significantly reduced grid size in $\mu$ is now required for the system under study to attain reliable results at low $T$ in comparison with that of the single-band Hubbard model.
In Fig. \ref{fig9} we show the isentropic lines in the $T-U$ plane, including the FM-ordered phases at different values of the lattice fillings: $n=0.85$, $n=1.2$, $n=1.6$, and $n=1.8$ (with fixed $V=22$ and $V_{ex}=21$). 

Note that in SU($N$)-symmetric mixtures an enhanced Pomeranchuk effect takes place. The Pomeranchuk effect was observed first in fermionic  $^3$He, where below a certain temperature the entropy of liquid $^3$He was found to be less than that of the same system in the solid state. Thus, heat absorption occurs when the solid phase is formed.
In experiments with ultracold gases in the lattice, the same physical mechanism was also observed \cite{Taie}, but only in the magnetically-unordered (PM) regime so far. It remains a highly-important experimental goal to cool down ultracold gases in optical lattices to the regime where substantial long-range magnetic correlations can develop \cite{Rey_review}. Hence, the Pomeranchuk effect constitutes a powerful and very useful mechanism from the point of view of many-body cooling. With a proper account of magnetic-ordering effects it was shown that the Pomeranchuk effect is significantly enhanced for the SU(3)-symmetric mixture in comparison with two-component fermionic systems in optical lattices \cite{Sotnikov2}. 

However, in Fig. \ref{fig9} it is clearly visible that in the case of strong asymmetry in the intra- and inter-orbital interaction strengths, the Pomeranchuck effect does not play a crucial role; that is, we do not observe that an increase of the intra-orbital interaction strength $U$ at constant entropy leads to a decrease in temperature, although at $n=0.85$ this behavior is  slightly visible. Note that a non-intuitive behavior of the isentropic lines is observed at $U\in[0,2]$ that, aside from limitations in the numerical analysis, can potentially be ascribed to the appearance of the superfluid instability in fermionic mixtures with a large asymmetry in interaction strengths \cite{Inaba}.

As for the entropy analysis in the vicinity of the FM-ordered phase, we observe a characteristic bending of the isentropic curves at the phase boundary that then follow the curvature of the magnetic phase transition lines (thick black lines in Fig. \ref{fig9}). Note that the same behavior is observed in the DMFT analysis of the AFM-ordered phases in the SU(2)- and SU(3)-symmetric mixtures at $n=1$ (quantum magnetism  based on the super exchange mechanism) \cite{Sotnikov2, Golubeva}. 

Note also that at strong coupling and from $n=0.85$ to $n\approx 1.5$, we observe that the isentropic curves are systematically shifted towards lower temperatures. However, at higher values of $n$, i.e., from $n\approx 1.6$ to $n=1.8$, the tendency is opposite and the isentropic curves are shifted towards higher values of $T$. This could be due to the fact that for higher values of $n$ (around half filling), ferromagnetic ordering is strongly suppressed by the antiferromagnetic exchange interactions and the behavior of the isentropic curves is similar to the one for which AFM magnetic ordering dominates \cite{Golubeva}.
At intermediate values of $n$ (the optimal regime for FM ordering), our results show that the FM-ordered phases in four-component mixtures can be approached and studied in detail in experiments with ultracold alkaline-earth-metal-like atoms, since the long-range ordered states are found to be stable at relatively high entropy values (up to $s_c\approx1.05$) that are within the reach of current cooling techniques.

\section{Conclusions and Outlook}\label{sec.5}

We have studied the magnetic phase diagram of the two-orbital Hubbard model in the context of recent experiments with ultracold gases of alkaline-earth-metal-like atoms in optical lattices, including an analog of Hund's coupling in solid-state materials.  Such systems can be realized nowadays by tuning the relative amplitude of spin-exchange processes between particular spin components in these mixtures. 
Finite-temperature phase diagrams were
obtained for the cases of fixed chemical potential and fixed density (lattice filling) by using the DMFT approach for the simple cubic lattice geometry. 

We found that the FM order is stabilized by the exchange interaction (Ising-type Hund's coupling) for a wide range of atomic densities away from half-filling. We determined the critical values of the exchange $V_{ex}$ for the appearance of FM-ordered states in the system at $T>0$. 
In the strong-coupling regime, we determined the optimal values of doping for observing FM instabilities in ultracold gases. Close to half-filling, the ferromagnetic ordering is found to be suppressed by the antiferromagnetic Heisenberg exchange, and the two-sublattice AFM-ordered state appears in the phase diagram. 
We also analyzed and determined the upper boundaries of the phase-separated region between AFM and FM ordering in the $T-n$ phase diagram. 
In the entropy analysis, we identified possible advantages of four-component mixtures of alkaline-earth-metal-like atoms in comparison with alkali-metal atoms for approaching quantum magnetism in optical lattices due to higher critical entropy values.

In this paper, we have restricted our analysis to the same filling within each orbital, i.e., $n_{e}=n_{g}$ (or, equivalently, $\Delta=0$), and equal values of intra-orbital interactions, $U_{gg}=U_{ee}=U$. However, the realization of state-dependent potentials for four-component ultracold atomic mixtures also provides possibilities to study from a new perspective the Kondo lattice model \cite{Zhang, Zhang-2}, which has been successfully applied to explain the physics of heavy-fermion compounds and other magnetic materials, such as nickel
and manganese perovskites \cite{ScazzaPhD}.
We did not study magnetic order in the presence of an external trapping potential, which could be an interesting extension of our analysis. 
It could also be of interest to investigate the influence of mass imbalance (different bandwidths) on the stability of magnetic phases by fixing the hopping amplitude of fermions in the $e$ band to be much smaller than in the $g$ band. Moreover, taking into account $t_{g}\neq t_{e}$ will also give an opportunity to study orbital-selective Mott-Hubbard transitions \cite{Jakobi} of ultracold atomic mixtures in optical lattices.

\vspace*{0.5cm}
\begin{acknowledgments}
\vspace*{-0.2cm}
The authors thank S. F\"{o}lling, A. Golubeva, J. Kune\v{s}, and F. Scazza for many fruitful discussions. We also thank W. Hofstetter for careful reading of the manuscript, valuable comments, and discussions. We gratefully acknowledge funding from the German Science Foundation DFG via Sonderforschungsbereich SFB/TR 49, resources provided by the high-performance computer cluster LOEWE-CSC, and the  funding received from the European Research Council (ERC) under the European Union's Horizon 2020 research and innovation program (Grant Agreement No. 646807-EXMAG, A.S.).
\end{acknowledgments}

\bibliographystyle{apsrev4-1}
\bibliography{Orbital_Magnetism_ACichy}

\begin{thebibliography}{61}%
\makeatletter
\providecommand \@ifxundefined [1]{%
 \@ifx{#1\undefined}
}%
\providecommand \@ifnum [1]{%
 \ifnum #1\expandafter \@firstoftwo
 \else \expandafter \@secondoftwo
 \fi
}%
\providecommand \@ifx [1]{%
 \ifx #1\expandafter \@firstoftwo
 \else \expandafter \@secondoftwo
 \fi
}%
\providecommand \natexlab [1]{#1}%
\providecommand \enquote  [1]{``#1''}%
\providecommand \bibnamefont  [1]{#1}%
\providecommand \bibfnamefont [1]{#1}%
\providecommand \citenamefont [1]{#1}%
\providecommand \href@noop [0]{\@secondoftwo}%
\providecommand \href [0]{\begingroup \@sanitize@url \@href}%
\providecommand \@href[1]{\@@startlink{#1}\@@href}%
\providecommand \@@href[1]{\endgroup#1\@@endlink}%
\providecommand \@sanitize@url [0]{\catcode `\\12\catcode `\$12\catcode
  `\&12\catcode `\#12\catcode `\^12\catcode `\_12\catcode `\%12\relax}%
\providecommand \@@startlink[1]{}%
\providecommand \@@endlink[0]{}%
\providecommand \url  [0]{\begingroup\@sanitize@url \@url }%
\providecommand \@url [1]{\endgroup\@href {#1}{\urlprefix }}%
\providecommand \urlprefix  [0]{URL }%
\providecommand \Eprint [0]{\href }%
\providecommand \doibase [0]{http://dx.doi.org/}%
\providecommand \selectlanguage [0]{\@gobble}%
\providecommand \bibinfo  [0]{\@secondoftwo}%
\providecommand \bibfield  [0]{\@secondoftwo}%
\providecommand \translation [1]{[#1]}%
\providecommand \BibitemOpen [0]{}%
\providecommand \bibitemStop [0]{}%
\providecommand \bibitemNoStop [0]{.\EOS\space}%
\providecommand \EOS [0]{\spacefactor3000\relax}%
\providecommand \BibitemShut  [1]{\csname bibitem#1\endcsname}%
\let\auto@bib@innerbib\@empty
\bibitem [{\citenamefont {Greiner}\ \emph {et~al.}(2002)\citenamefont
  {Greiner}, \citenamefont {Mandel}, \citenamefont {Esslinger}, \citenamefont
  {H\"ansch},\ and\ \citenamefont {Bloch}}]{greiner}%
  \BibitemOpen
  \bibfield  {author} {\bibinfo {author} {\bibfnamefont {M.}~\bibnamefont
  {Greiner}}, \bibinfo {author} {\bibfnamefont {O.}~\bibnamefont {Mandel}},
  \bibinfo {author} {\bibfnamefont {T.}~\bibnamefont {Esslinger}}, \bibinfo
  {author} {\bibfnamefont {T.~W.}\ \bibnamefont {H\"ansch}}, \ and\ \bibinfo
  {author} {\bibfnamefont {I.}~\bibnamefont {Bloch}},\ }\href {\doibase
  10.1038/415039a} {\bibfield  {journal} {\bibinfo  {journal} {Nature}\
  }\textbf {\bibinfo {volume} {415}},\ \bibinfo {pages} {39} (\bibinfo {year}
  {2002})}\BibitemShut {NoStop}%
\bibitem [{\citenamefont {Chin}\ \emph {et~al.}(2004)\citenamefont {Chin},
  \citenamefont {Bartenstein}, \citenamefont {Altmeyer}, \citenamefont {Riedl},
  \citenamefont {Jochim}, \citenamefont {Denschlag},\ and\ \citenamefont
  {Grimm}}]{chin}%
  \BibitemOpen
  \bibfield  {author} {\bibinfo {author} {\bibfnamefont {C.}~\bibnamefont
  {Chin}}, \bibinfo {author} {\bibfnamefont {M.}~\bibnamefont {Bartenstein}},
  \bibinfo {author} {\bibfnamefont {A.}~\bibnamefont {Altmeyer}}, \bibinfo
  {author} {\bibfnamefont {S.}~\bibnamefont {Riedl}}, \bibinfo {author}
  {\bibfnamefont {S.}~\bibnamefont {Jochim}}, \bibinfo {author} {\bibfnamefont
  {J.~H.}\ \bibnamefont {Denschlag}}, \ and\ \bibinfo {author} {\bibfnamefont
  {R.}~\bibnamefont {Grimm}},\ }\href {\doibase 10.1126/science.1100818}
  {\bibfield  {journal} {\bibinfo  {journal} {Science}\ }\textbf {\bibinfo
  {volume} {305}},\ \bibinfo {pages} {1128} (\bibinfo {year}
  {2004})}\BibitemShut {NoStop}%
\bibitem [{\citenamefont {K\"ohl}\ \emph {et~al.}(2005)\citenamefont {K\"ohl},
  \citenamefont {Moritz}, \citenamefont {St\"oferle}, \citenamefont
  {G\"unter},\ and\ \citenamefont {Esslinger}}]{koln}%
  \BibitemOpen
  \bibfield  {author} {\bibinfo {author} {\bibfnamefont {M.}~\bibnamefont
  {K\"ohl}}, \bibinfo {author} {\bibfnamefont {H.}~\bibnamefont {Moritz}},
  \bibinfo {author} {\bibfnamefont {T.}~\bibnamefont {St\"oferle}}, \bibinfo
  {author} {\bibfnamefont {K.}~\bibnamefont {G\"unter}}, \ and\ \bibinfo
  {author} {\bibfnamefont {T.}~\bibnamefont {Esslinger}},\ }\href {\doibase
  10.1103/PhysRevLett.94.080403} {\bibfield  {journal} {\bibinfo  {journal}
  {Phys. Rev. Lett.}\ }\textbf {\bibinfo {volume} {94}},\ \bibinfo {pages}
  {080403} (\bibinfo {year} {2005})}\BibitemShut {NoStop}%
\bibitem [{\citenamefont {St\"oferle}\ \emph {et~al.}(2006)\citenamefont
  {St\"oferle}, \citenamefont {Moritz}, \citenamefont {G\"unter}, \citenamefont
  {K\"ohl},\ and\ \citenamefont {Esslinger}}]{stoferle}%
  \BibitemOpen
  \bibfield  {author} {\bibinfo {author} {\bibfnamefont {T.}~\bibnamefont
  {St\"oferle}}, \bibinfo {author} {\bibfnamefont {H.}~\bibnamefont {Moritz}},
  \bibinfo {author} {\bibfnamefont {K.}~\bibnamefont {G\"unter}}, \bibinfo
  {author} {\bibfnamefont {M.}~\bibnamefont {K\"ohl}}, \ and\ \bibinfo {author}
  {\bibfnamefont {T.}~\bibnamefont {Esslinger}},\ }\href {\doibase
  10.1103/PhysRevLett.96.030401} {\bibfield  {journal} {\bibinfo  {journal}
  {Phys. Rev. Lett.}\ }\textbf {\bibinfo {volume} {96}},\ \bibinfo {pages}
  {030401} (\bibinfo {year} {2006})}\BibitemShut {NoStop}%
\bibitem [{\citenamefont {Bloch}\ \emph {et~al.}(2008)\citenamefont {Bloch},
  \citenamefont {Dalibard},\ and\ \citenamefont {Zwerger}}]{Blo2008RMP}%
  \BibitemOpen
  \bibfield  {author} {\bibinfo {author} {\bibfnamefont {I.}~\bibnamefont
  {Bloch}}, \bibinfo {author} {\bibfnamefont {J.}~\bibnamefont {Dalibard}}, \
  and\ \bibinfo {author} {\bibfnamefont {W.}~\bibnamefont {Zwerger}},\ }\href
  {\doibase 10.1103/RevModPhys.80.885} {\bibfield  {journal} {\bibinfo
  {journal} {Rev. Mod. Phys.}\ }\textbf {\bibinfo {volume} {80}},\ \bibinfo
  {pages} {885} (\bibinfo {year} {2008})}\BibitemShut {NoStop}%
\bibitem [{\citenamefont {J\"ordens}\ \emph {et~al.}(2010)\citenamefont
  {J\"ordens}, \citenamefont {Tarruell}, \citenamefont {Greif}, \citenamefont
  {Uehlinger}, \citenamefont {Strohmaier}, \citenamefont {Moritz},
  \citenamefont {Esslinger}, \citenamefont {De~Leo}, \citenamefont {Kollath},
  \citenamefont {Georges}, \citenamefont {Scarola}, \citenamefont {Pollet},
  \citenamefont {Burovski}, \citenamefont {Kozik},\ and\ \citenamefont
  {Troyer}}]{Jor2010PRL}%
  \BibitemOpen
  \bibfield  {author} {\bibinfo {author} {\bibfnamefont {R.}~\bibnamefont
  {J\"ordens}}, \bibinfo {author} {\bibfnamefont {L.}~\bibnamefont {Tarruell}},
  \bibinfo {author} {\bibfnamefont {D.}~\bibnamefont {Greif}}, \bibinfo
  {author} {\bibfnamefont {T.}~\bibnamefont {Uehlinger}}, \bibinfo {author}
  {\bibfnamefont {N.}~\bibnamefont {Strohmaier}}, \bibinfo {author}
  {\bibfnamefont {H.}~\bibnamefont {Moritz}}, \bibinfo {author} {\bibfnamefont
  {T.}~\bibnamefont {Esslinger}}, \bibinfo {author} {\bibfnamefont
  {L.}~\bibnamefont {De~Leo}}, \bibinfo {author} {\bibfnamefont
  {C.}~\bibnamefont {Kollath}}, \bibinfo {author} {\bibfnamefont
  {A.}~\bibnamefont {Georges}}, \bibinfo {author} {\bibfnamefont
  {V.}~\bibnamefont {Scarola}}, \bibinfo {author} {\bibfnamefont
  {L.}~\bibnamefont {Pollet}}, \bibinfo {author} {\bibfnamefont
  {E.}~\bibnamefont {Burovski}}, \bibinfo {author} {\bibfnamefont
  {E.}~\bibnamefont {Kozik}}, \ and\ \bibinfo {author} {\bibfnamefont
  {M.}~\bibnamefont {Troyer}},\ }\href {\doibase
  10.1103/PhysRevLett.104.180401} {\bibfield  {journal} {\bibinfo  {journal}
  {Phys. Rev. Lett.}\ }\textbf {\bibinfo {volume} {104}},\ \bibinfo {pages}
  {180401} (\bibinfo {year} {2010})}\BibitemShut {NoStop}%
\bibitem [{\citenamefont {Schneider}(2010)}]{SchneiderThesis}%
  \BibitemOpen
  \bibfield  {author} {\bibinfo {author} {\bibfnamefont {U.}~\bibnamefont
  {Schneider}},\ }\emph {\bibinfo {title} {Interacting Fermionic Atoms in
  Optical Lattices -- A Quantum Simulator for Condensed Matter Physics}},\
  \href@noop {} {Ph.D. thesis},\ \bibinfo  {school} {Johannes
  Gutenberg-Universit\"at in Mainz} (\bibinfo {year} {2010})\BibitemShut
  {NoStop}%
\bibitem [{\citenamefont {Fulde}\ and\ \citenamefont {Ferrell}(1964)}]{fulde}%
  \BibitemOpen
  \bibfield  {author} {\bibinfo {author} {\bibfnamefont {P.}~\bibnamefont
  {Fulde}}\ and\ \bibinfo {author} {\bibfnamefont {R.~A.}\ \bibnamefont
  {Ferrell}},\ }\href@noop {} {\bibfield  {journal} {\bibinfo  {journal} {Phys.
  Rev.}\ }\textbf {\bibinfo {volume} {135}},\ \bibinfo {pages} {A550} (\bibinfo
  {year} {1964})}\BibitemShut {NoStop}%
\bibitem [{\citenamefont {Larkin}\ and\ \citenamefont
  {Ovchinnikov}(1964)}]{Larkin}%
  \BibitemOpen
  \bibfield  {author} {\bibinfo {author} {\bibfnamefont {A.~I.}\ \bibnamefont
  {Larkin}}\ and\ \bibinfo {author} {\bibfnamefont {Y.~N.}\ \bibnamefont
  {Ovchinnikov}},\ }\href@noop {} {\bibfield  {journal} {\bibinfo  {journal}
  {Zh. Eksp. Teor. Fiz.}\ }\textbf {\bibinfo {volume} {47}},\ \bibinfo {pages}
  {1136} (\bibinfo {year} {1964})}\BibitemShut {NoStop}%
\bibitem [{\citenamefont {Bianchi}\ \emph {et~al.}(2003)\citenamefont
  {Bianchi}, \citenamefont {Movshovich}, \citenamefont {Capan}, \citenamefont
  {Pagliuso},\ and\ \citenamefont {Sarrao}}]{Bianchi}%
  \BibitemOpen
  \bibfield  {author} {\bibinfo {author} {\bibfnamefont {A.}~\bibnamefont
  {Bianchi}}, \bibinfo {author} {\bibfnamefont {R.}~\bibnamefont {Movshovich}},
  \bibinfo {author} {\bibfnamefont {C.}~\bibnamefont {Capan}}, \bibinfo
  {author} {\bibfnamefont {P.~G.}\ \bibnamefont {Pagliuso}}, \ and\ \bibinfo
  {author} {\bibfnamefont {J.~L.}\ \bibnamefont {Sarrao}},\ }\href@noop {}
  {\bibfield  {journal} {\bibinfo  {journal} {Phys. Rev. Lett.}\ }\textbf
  {\bibinfo {volume} {91}},\ \bibinfo {pages} {187004} (\bibinfo {year}
  {2003})}\BibitemShut {NoStop}%
\bibitem [{\citenamefont {Casalbuoni}\ and\ \citenamefont
  {Nardulli}(2004)}]{Nardulli}%
  \BibitemOpen
  \bibfield  {author} {\bibinfo {author} {\bibfnamefont {R.}~\bibnamefont
  {Casalbuoni}}\ and\ \bibinfo {author} {\bibfnamefont {G.}~\bibnamefont
  {Nardulli}},\ }\href@noop {} {\bibfield  {journal} {\bibinfo  {journal} {Rev.
  Mod. Phys}\ }\textbf {\bibinfo {volume} {76}},\ \bibinfo {pages} {263}
  (\bibinfo {year} {2004})}\BibitemShut {NoStop}%
\bibitem [{\citenamefont {for a review~see: Y.~Matsuda}\ and\ \citenamefont
  {Shimahara}(2007)}]{Matsuda}%
  \BibitemOpen
  \bibfield  {author} {\bibinfo {author} {\bibnamefont {for a review~see:
  Y.~Matsuda}}\ and\ \bibinfo {author} {\bibfnamefont {H.}~\bibnamefont
  {Shimahara}},\ }\href@noop {} {\bibfield  {journal} {\bibinfo  {journal} {J.
  Phys. Soc. Jpn.}\ }\textbf {\bibinfo {volume} {76}},\ \bibinfo {pages}
  {051005} (\bibinfo {year} {2007})}\BibitemShut {NoStop}%
\bibitem [{\citenamefont {Koponen}\ \emph {et~al.}(2008)\citenamefont
  {Koponen}, \citenamefont {Paananen}, \citenamefont {Martikainen},
  \citenamefont {Bakhtiari},\ and\ \citenamefont {Torma}}]{torma}%
  \BibitemOpen
  \bibfield  {author} {\bibinfo {author} {\bibfnamefont {T.~K.}\ \bibnamefont
  {Koponen}}, \bibinfo {author} {\bibfnamefont {T.}~\bibnamefont {Paananen}},
  \bibinfo {author} {\bibfnamefont {J.-P.}\ \bibnamefont {Martikainen}},
  \bibinfo {author} {\bibfnamefont {M.~R.}\ \bibnamefont {Bakhtiari}}, \ and\
  \bibinfo {author} {\bibfnamefont {P.}~\bibnamefont {Torma}},\ }\href@noop {}
  {\bibfield  {journal} {\bibinfo  {journal} {New Journal of Physics}\ }\textbf
  {\bibinfo {volume} {10}},\ \bibinfo {pages} {045104} (\bibinfo {year}
  {2008})}\BibitemShut {NoStop}%
\bibitem [{\citenamefont {Koponen}\ \emph {et~al.}(2007)\citenamefont
  {Koponen}, \citenamefont {Paananen}, \citenamefont {Martikainen},\ and\
  \citenamefont {Torma}}]{koponen}%
  \BibitemOpen
  \bibfield  {author} {\bibinfo {author} {\bibfnamefont {T.~K.}\ \bibnamefont
  {Koponen}}, \bibinfo {author} {\bibfnamefont {T.}~\bibnamefont {Paananen}},
  \bibinfo {author} {\bibfnamefont {J.-P.}\ \bibnamefont {Martikainen}}, \ and\
  \bibinfo {author} {\bibfnamefont {P.}~\bibnamefont {Torma}},\ }\href@noop {}
  {\bibfield  {journal} {\bibinfo  {journal} {Phys. Rev. Lett.}\ }\textbf
  {\bibinfo {volume} {99}},\ \bibinfo {pages} {120403} (\bibinfo {year}
  {2007})}\BibitemShut {NoStop}%
\bibitem [{\citenamefont {Mierzejewski}\ \emph {et~al.}(2009)\citenamefont
  {Mierzejewski}, \citenamefont {Ptok},\ and\ \citenamefont
  {Maska}}]{Mierzejewski}%
  \BibitemOpen
  \bibfield  {author} {\bibinfo {author} {\bibfnamefont {M.}~\bibnamefont
  {Mierzejewski}}, \bibinfo {author} {\bibfnamefont {A.}~\bibnamefont {Ptok}},
  \ and\ \bibinfo {author} {\bibfnamefont {M.~M.}\ \bibnamefont {Maska}},\
  }\href@noop {} {\bibfield  {journal} {\bibinfo  {journal} {Phys. Rev. B}\
  }\textbf {\bibinfo {volume} {80}},\ \bibinfo {pages} {174525} (\bibinfo
  {year} {2009})}\BibitemShut {NoStop}%
\bibitem [{\citenamefont {Loh}\ and\ \citenamefont {Trivedi}(2008)}]{loh}%
  \BibitemOpen
  \bibfield  {author} {\bibinfo {author} {\bibfnamefont {Y.~L.}\ \bibnamefont
  {Loh}}\ and\ \bibinfo {author} {\bibfnamefont {N.}~\bibnamefont {Trivedi}},\
  }\href@noop {} {\bibfield  {journal} {\bibinfo  {journal} {Phys. Rev. Lett.}\
  }\textbf {\bibinfo {volume} {104}},\ \bibinfo {pages} {165302} (\bibinfo
  {year} {2008})}\BibitemShut {NoStop}%
\bibitem [{\citenamefont {Micnas}\ \emph {et~al.}(1990)\citenamefont {Micnas},
  \citenamefont {Ranninger},\ and\ \citenamefont
  {Robaszkiewicz}}]{MicnasModern}%
  \BibitemOpen
  \bibfield  {author} {\bibinfo {author} {\bibfnamefont {R.}~\bibnamefont
  {Micnas}}, \bibinfo {author} {\bibfnamefont {J.}~\bibnamefont {Ranninger}}, \
  and\ \bibinfo {author} {\bibfnamefont {S.}~\bibnamefont {Robaszkiewicz}},\
  }\href@noop {} {\bibfield  {journal} {\bibinfo  {journal} {Rev. Mod. Phys.}\
  }\textbf {\bibinfo {volume} {62}},\ \bibinfo {pages} {113} (\bibinfo {year}
  {1990})}\BibitemShut {NoStop}%
\bibitem [{\citenamefont {Zwierlein}\ \emph {et~al.}(2005)\citenamefont
  {Zwierlein}, \citenamefont {Abo-Shaeer}, \citenamefont {Schirotzek},
  \citenamefont {Schunck},\ and\ \citenamefont {Ketterle}}]{ketterle}%
  \BibitemOpen
  \bibfield  {author} {\bibinfo {author} {\bibfnamefont {M.~W.}\ \bibnamefont
  {Zwierlein}}, \bibinfo {author} {\bibfnamefont {J.}~\bibnamefont
  {Abo-Shaeer}}, \bibinfo {author} {\bibfnamefont {A.}~\bibnamefont
  {Schirotzek}}, \bibinfo {author} {\bibfnamefont {C.}~\bibnamefont {Schunck}},
  \ and\ \bibinfo {author} {\bibfnamefont {W.}~\bibnamefont {Ketterle}},\
  }\href@noop {} {\bibfield  {journal} {\bibinfo  {journal} {Nature}\ }\textbf
  {\bibinfo {volume} {435}},\ \bibinfo {pages} {1047} (\bibinfo {year}
  {2005})}\BibitemShut {NoStop}%
\bibitem [{\citenamefont {Zwierlein}\ \emph {et~al.}(2006)\citenamefont
  {Zwierlein}, \citenamefont {Schirotzek}, \citenamefont {Schunck},\ and\
  \citenamefont {Ketterle}}]{ketterle2}%
  \BibitemOpen
  \bibfield  {author} {\bibinfo {author} {\bibfnamefont {M.~W.}\ \bibnamefont
  {Zwierlein}}, \bibinfo {author} {\bibfnamefont {A.}~\bibnamefont
  {Schirotzek}}, \bibinfo {author} {\bibfnamefont {C.~H.}\ \bibnamefont
  {Schunck}}, \ and\ \bibinfo {author} {\bibfnamefont {W.}~\bibnamefont
  {Ketterle}},\ }\href@noop {} {\bibfield  {journal} {\bibinfo  {journal}
  {Science}\ }\textbf {\bibinfo {volume} {311}},\ \bibinfo {pages} {492}
  (\bibinfo {year} {2006})}\BibitemShut {NoStop}%
\bibitem [{\citenamefont {Titvinidze}\ \emph {et~al.}(2009)\citenamefont
  {Titvinidze}, \citenamefont {Snoek},\ and\ \citenamefont
  {Hofstetter}}]{titvinidze}%
  \BibitemOpen
  \bibfield  {author} {\bibinfo {author} {\bibfnamefont {I.}~\bibnamefont
  {Titvinidze}}, \bibinfo {author} {\bibfnamefont {M.}~\bibnamefont {Snoek}}, \
  and\ \bibinfo {author} {\bibfnamefont {W.}~\bibnamefont {Hofstetter}},\
  }\href@noop {} {\bibfield  {journal} {\bibinfo  {journal} {Phys. Rev. B}\
  }\textbf {\bibinfo {volume} {79}},\ \bibinfo {pages} {144506} (\bibinfo
  {year} {2009})}\BibitemShut {NoStop}%
\bibitem [{\citenamefont {Snoek}\ \emph {et~al.}(2010)\citenamefont {Snoek},
  \citenamefont {Titvinidze}, \citenamefont {Bloch},\ and\ \citenamefont
  {Hofstetter}}]{titvinidze2}%
  \BibitemOpen
  \bibfield  {author} {\bibinfo {author} {\bibfnamefont {M.}~\bibnamefont
  {Snoek}}, \bibinfo {author} {\bibfnamefont {I.}~\bibnamefont {Titvinidze}},
  \bibinfo {author} {\bibfnamefont {I.}~\bibnamefont {Bloch}}, \ and\ \bibinfo
  {author} {\bibfnamefont {W.}~\bibnamefont {Hofstetter}},\ }\href@noop {}
  {\bibfield  {journal} {\bibinfo  {journal} {Phys. Rev. Lett.}\ }\textbf
  {\bibinfo {volume} {106}},\ \bibinfo {pages} {155301} (\bibinfo {year}
  {2010})}\BibitemShut {NoStop}%
\bibitem [{\citenamefont {Honerkamp}\ and\ \citenamefont
  {Hofstetter}(2004)}]{Hofstetter}%
  \BibitemOpen
  \bibfield  {author} {\bibinfo {author} {\bibfnamefont {C.}~\bibnamefont
  {Honerkamp}}\ and\ \bibinfo {author} {\bibfnamefont {W.}~\bibnamefont
  {Hofstetter}},\ }\href {\doibase 10.1103/PhysRevLett.92.170403} {\bibfield
  {journal} {\bibinfo  {journal} {Phys. Rev. Lett}\ }\textbf {\bibinfo {volume}
  {92}},\ \bibinfo {pages} {170403} (\bibinfo {year} {2004})}\BibitemShut
  {NoStop}%
\bibitem [{\citenamefont {Privitera}\ \emph {et~al.}(2011)\citenamefont
  {Privitera}, \citenamefont {Titvinidze}, \citenamefont {Chang}, \citenamefont
  {Diehl}, \citenamefont {Daley},\ and\ \citenamefont
  {Hofstetter}}]{privitera}%
  \BibitemOpen
  \bibfield  {author} {\bibinfo {author} {\bibfnamefont {A.}~\bibnamefont
  {Privitera}}, \bibinfo {author} {\bibfnamefont {I.}~\bibnamefont
  {Titvinidze}}, \bibinfo {author} {\bibfnamefont {S.-Y.}\ \bibnamefont
  {Chang}}, \bibinfo {author} {\bibfnamefont {S.}~\bibnamefont {Diehl}},
  \bibinfo {author} {\bibfnamefont {A.~J.}\ \bibnamefont {Daley}}, \ and\
  \bibinfo {author} {\bibfnamefont {W.}~\bibnamefont {Hofstetter}},\
  }\href@noop {} {\bibfield  {journal} {\bibinfo  {journal} {Phys. Rev. A}\
  }\textbf {\bibinfo {volume} {84}},\ \bibinfo {pages} {021601(R)} (\bibinfo
  {year} {2011})}\BibitemShut {NoStop}%
\bibitem [{\citenamefont {Cazalilla}\ and\ \citenamefont
  {Rey}(2014)}]{Rey_review}%
  \BibitemOpen
  \bibfield  {author} {\bibinfo {author} {\bibfnamefont {M.~A.}\ \bibnamefont
  {Cazalilla}}\ and\ \bibinfo {author} {\bibfnamefont {A.~M.}\ \bibnamefont
  {Rey}},\ }\href {\doibase 10.1088/0034-4885/77/12/124401} {\bibfield
  {journal} {\bibinfo  {journal} {Rep. Prog. Phys.}\ }\textbf {\bibinfo
  {volume} {77}},\ \bibinfo {pages} {124401} (\bibinfo {year}
  {2014})}\BibitemShut {NoStop}%
\bibitem [{\citenamefont {Greif}\ \emph {et~al.}(2013)\citenamefont {Greif},
  \citenamefont {Uehlinger}, \citenamefont {Jotzu}, \citenamefont {Tarruell},\
  and\ \citenamefont {Esslinger}}]{Greif}%
  \BibitemOpen
  \bibfield  {author} {\bibinfo {author} {\bibfnamefont {D.}~\bibnamefont
  {Greif}}, \bibinfo {author} {\bibfnamefont {T.}~\bibnamefont {Uehlinger}},
  \bibinfo {author} {\bibfnamefont {G.}~\bibnamefont {Jotzu}}, \bibinfo
  {author} {\bibfnamefont {L.}~\bibnamefont {Tarruell}}, \ and\ \bibinfo
  {author} {\bibfnamefont {T.}~\bibnamefont {Esslinger}},\ }\href {\doibase
  10.1126/science.1236362} {\bibfield  {journal} {\bibinfo  {journal}
  {Science}\ }\textbf {\bibinfo {volume} {340}},\ \bibinfo {pages} {1307}
  (\bibinfo {year} {2013})}\BibitemShut {NoStop}%
\bibitem [{\citenamefont {Hart}\ \emph {et~al.}(2015)\citenamefont {Hart},
  \citenamefont {Duarte}, \citenamefont {Yang}, \citenamefont {Liu},
  \citenamefont {Paiva}, \citenamefont {Khatami}, \citenamefont {Scalettar},
  \citenamefont {Trivedi}, \citenamefont {Huse},\ and\ \citenamefont
  {Hulet}}]{Har2015Nat}%
  \BibitemOpen
  \bibfield  {author} {\bibinfo {author} {\bibfnamefont {R.~A.}\ \bibnamefont
  {Hart}}, \bibinfo {author} {\bibfnamefont {P.~M.}\ \bibnamefont {Duarte}},
  \bibinfo {author} {\bibfnamefont {T.-L.}\ \bibnamefont {Yang}}, \bibinfo
  {author} {\bibfnamefont {X.}~\bibnamefont {Liu}}, \bibinfo {author}
  {\bibfnamefont {T.}~\bibnamefont {Paiva}}, \bibinfo {author} {\bibfnamefont
  {E.}~\bibnamefont {Khatami}}, \bibinfo {author} {\bibfnamefont {R.~T.}\
  \bibnamefont {Scalettar}}, \bibinfo {author} {\bibfnamefont {N.}~\bibnamefont
  {Trivedi}}, \bibinfo {author} {\bibfnamefont {D.~A.}\ \bibnamefont {Huse}}, \
  and\ \bibinfo {author} {\bibfnamefont {R.~G.}\ \bibnamefont {Hulet}},\ }\href
  {http://dx.doi.org/10.1038/nature14223} {\bibfield  {journal} {\bibinfo
  {journal} {Nature}\ }\textbf {\bibinfo {volume} {519}},\ \bibinfo {pages}
  {211} (\bibinfo {year} {2015})}\BibitemShut {NoStop}%
\bibitem [{\citenamefont {Taie}\ \emph {et~al.}(2010)\citenamefont {Taie},
  \citenamefont {Takasu}, \citenamefont {Sugawa}, \citenamefont {Yamazaki},
  \citenamefont {Tsujimoto}, \citenamefont {Murakami},\ and\ \citenamefont
  {Takahashi}}]{Takahashi}%
  \BibitemOpen
  \bibfield  {author} {\bibinfo {author} {\bibfnamefont {S.}~\bibnamefont
  {Taie}}, \bibinfo {author} {\bibfnamefont {Y.}~\bibnamefont {Takasu}},
  \bibinfo {author} {\bibfnamefont {S.}~\bibnamefont {Sugawa}}, \bibinfo
  {author} {\bibfnamefont {R.}~\bibnamefont {Yamazaki}}, \bibinfo {author}
  {\bibfnamefont {T.}~\bibnamefont {Tsujimoto}}, \bibinfo {author}
  {\bibfnamefont {R.}~\bibnamefont {Murakami}}, \ and\ \bibinfo {author}
  {\bibfnamefont {Y.}~\bibnamefont {Takahashi}},\ }\href {\doibase
  10.1103/PhysRevLett.105.190401} {\bibfield  {journal} {\bibinfo  {journal}
  {Phys. Rev. Lett.}\ }\textbf {\bibinfo {volume} {105}},\ \bibinfo {pages}
  {190401} (\bibinfo {year} {2010})}\BibitemShut {NoStop}%
\bibitem [{\citenamefont {Taie}\ \emph {et~al.}(2012)\citenamefont {Taie},
  \citenamefont {Yamazaki}, \citenamefont {Sugawa},\ and\ \citenamefont
  {Takahashi}}]{Taie}%
  \BibitemOpen
  \bibfield  {author} {\bibinfo {author} {\bibfnamefont {S.}~\bibnamefont
  {Taie}}, \bibinfo {author} {\bibfnamefont {R.}~\bibnamefont {Yamazaki}},
  \bibinfo {author} {\bibfnamefont {S.}~\bibnamefont {Sugawa}}, \ and\ \bibinfo
  {author} {\bibfnamefont {Y.}~\bibnamefont {Takahashi}},\ }\href {\doibase
  10.1038/nphys2430} {\bibfield  {journal} {\bibinfo  {journal} {Nature
  Physics}\ }\textbf {\bibinfo {volume} {8}},\ \bibinfo {pages} {825} (\bibinfo
  {year} {2012})}\BibitemShut {NoStop}%
\bibitem [{\citenamefont {Thobe}(2014)}]{Thobe}%
  \BibitemOpen
  \bibfield  {author} {\bibinfo {author} {\bibfnamefont {A.}~\bibnamefont
  {Thobe}},\ }\emph {\bibinfo {title} {Ultracold Yb Gases with Control over
  Spin and Orbital Degrees of Freedom}},\ \href@noop {} {Ph.D. thesis},\
  \bibinfo  {school} {Universit\"at Hamburg} (\bibinfo {year}
  {2014})\BibitemShut {NoStop}%
\bibitem [{\citenamefont {Cappellini}\ \emph {et~al.}(2014)\citenamefont
  {Cappellini}, \citenamefont {Mancini}, \citenamefont {Pagano}, \citenamefont
  {Lombardi}, \citenamefont {Livi}, \citenamefont {Siciliani~de Cumis},
  \citenamefont {Cancio}, \citenamefont {Pizzocaro}, \citenamefont {Calonico},
  \citenamefont {Levi}, \citenamefont {Sias}, \citenamefont {Catani},
  \citenamefont {Inguscio},\ and\ \citenamefont {Fallani}}]{Cappellini}%
  \BibitemOpen
  \bibfield  {author} {\bibinfo {author} {\bibfnamefont {G.}~\bibnamefont
  {Cappellini}}, \bibinfo {author} {\bibfnamefont {M.}~\bibnamefont {Mancini}},
  \bibinfo {author} {\bibfnamefont {G.}~\bibnamefont {Pagano}}, \bibinfo
  {author} {\bibfnamefont {P.}~\bibnamefont {Lombardi}}, \bibinfo {author}
  {\bibfnamefont {L.}~\bibnamefont {Livi}}, \bibinfo {author} {\bibfnamefont
  {M.}~\bibnamefont {Siciliani~de Cumis}}, \bibinfo {author} {\bibfnamefont
  {P.}~\bibnamefont {Cancio}}, \bibinfo {author} {\bibfnamefont
  {M.}~\bibnamefont {Pizzocaro}}, \bibinfo {author} {\bibfnamefont
  {D.}~\bibnamefont {Calonico}}, \bibinfo {author} {\bibfnamefont
  {F.}~\bibnamefont {Levi}}, \bibinfo {author} {\bibfnamefont {C.}~\bibnamefont
  {Sias}}, \bibinfo {author} {\bibfnamefont {J.}~\bibnamefont {Catani}},
  \bibinfo {author} {\bibfnamefont {M.}~\bibnamefont {Inguscio}}, \ and\
  \bibinfo {author} {\bibfnamefont {L.}~\bibnamefont {Fallani}},\ }\href
  {\doibase 10.1103/PhysRevLett.113.120402} {\bibfield  {journal} {\bibinfo
  {journal} {Phys. Rev. Lett.}\ }\textbf {\bibinfo {volume} {113}},\ \bibinfo
  {pages} {120402} (\bibinfo {year} {2014})}\BibitemShut {NoStop}%
\bibitem [{\citenamefont {Pagano}\ \emph {et~al.}(2014)\citenamefont {Pagano},
  \citenamefont {Cappellini}, \citenamefont {Mancini}, \citenamefont
  {Lombardi}, \citenamefont {Sch\"afer}, \citenamefont {Hu}, \citenamefont
  {Liu}, \citenamefont {Catani}, \citenamefont {Sias},\ and\ \citenamefont
  {Inguscio}}]{Pagano}%
  \BibitemOpen
  \bibfield  {author} {\bibinfo {author} {\bibfnamefont {G.}~\bibnamefont
  {Pagano}}, \bibinfo {author} {\bibfnamefont {G.}~\bibnamefont {Cappellini}},
  \bibinfo {author} {\bibfnamefont {M.}~\bibnamefont {Mancini}}, \bibinfo
  {author} {\bibfnamefont {P.}~\bibnamefont {Lombardi}}, \bibinfo {author}
  {\bibfnamefont {F.}~\bibnamefont {Sch\"afer}}, \bibinfo {author}
  {\bibfnamefont {H.}~\bibnamefont {Hu}}, \bibinfo {author} {\bibfnamefont
  {X.-J.}\ \bibnamefont {Liu}}, \bibinfo {author} {\bibfnamefont
  {J.}~\bibnamefont {Catani}}, \bibinfo {author} {\bibfnamefont
  {C.}~\bibnamefont {Sias}}, \ and\ \bibinfo {author} {\bibfnamefont
  {M.}~\bibnamefont {Inguscio}},\ }\href {\doibase 10.1038/nphys2878}
  {\bibfield  {journal} {\bibinfo  {journal} {Nature Physics}\ }\textbf
  {\bibinfo {volume} {10}},\ \bibinfo {pages} {198} (\bibinfo {year}
  {2014})}\BibitemShut {NoStop}%
\bibitem [{\citenamefont {{Scazza}}\ \emph {et~al.}(2014)\citenamefont
  {{Scazza}}, \citenamefont {{Hofrichter}}, \citenamefont {{H{\"o}fer}},
  \citenamefont {{de Groot}}, \citenamefont {{Bloch}},\ and\ \citenamefont
  {{F{\"o}lling}}}]{Scazza}%
  \BibitemOpen
  \bibfield  {author} {\bibinfo {author} {\bibfnamefont {F.}~\bibnamefont
  {{Scazza}}}, \bibinfo {author} {\bibfnamefont {C.}~\bibnamefont
  {{Hofrichter}}}, \bibinfo {author} {\bibfnamefont {M.}~\bibnamefont
  {{H{\"o}fer}}}, \bibinfo {author} {\bibfnamefont {P.~C.}\ \bibnamefont {{de
  Groot}}}, \bibinfo {author} {\bibfnamefont {I.}~\bibnamefont {{Bloch}}}, \
  and\ \bibinfo {author} {\bibfnamefont {S.}~\bibnamefont {{F{\"o}lling}}},\
  }\href {\doibase 10.1038/nphys3061} {\bibfield  {journal} {\bibinfo
  {journal} {Nature Physics}\ }\textbf {\bibinfo {volume} {10}},\ \bibinfo
  {pages} {779} (\bibinfo {year} {2014})}\BibitemShut {NoStop}%
\bibitem [{\citenamefont {Gorshkov}\ \emph {et~al.}(2010)\citenamefont
  {Gorshkov}, \citenamefont {Hermele}, \citenamefont {Gurarie}, \citenamefont
  {Xu}, \citenamefont {Julienne}, \citenamefont {Ye}, \citenamefont {Zoller},
  \citenamefont {Demler}, \citenamefont {Lukin},\ and\ \citenamefont
  {Rey}}]{Gorshkov}%
  \BibitemOpen
  \bibfield  {author} {\bibinfo {author} {\bibfnamefont {A.~V.}\ \bibnamefont
  {Gorshkov}}, \bibinfo {author} {\bibfnamefont {M.}~\bibnamefont {Hermele}},
  \bibinfo {author} {\bibfnamefont {V.}~\bibnamefont {Gurarie}}, \bibinfo
  {author} {\bibfnamefont {C.}~\bibnamefont {Xu}}, \bibinfo {author}
  {\bibfnamefont {P.~S.}\ \bibnamefont {Julienne}}, \bibinfo {author}
  {\bibfnamefont {J.}~\bibnamefont {Ye}}, \bibinfo {author} {\bibfnamefont
  {P.}~\bibnamefont {Zoller}}, \bibinfo {author} {\bibfnamefont
  {E.}~\bibnamefont {Demler}}, \bibinfo {author} {\bibfnamefont {M.~D.}\
  \bibnamefont {Lukin}}, \ and\ \bibinfo {author} {\bibfnamefont {A.~M.}\
  \bibnamefont {Rey}},\ }\href {\doibase 10.1038/nphys1535} {\bibfield
  {journal} {\bibinfo  {journal} {Nature Physics}\ }\textbf {\bibinfo {volume}
  {6}},\ \bibinfo {pages} {289} (\bibinfo {year} {2010})}\BibitemShut {NoStop}%
\bibitem [{\citenamefont {Foss-Feig}\ \emph
  {et~al.}(2010{\natexlab{a}})\citenamefont {Foss-Feig}, \citenamefont
  {Hermele}, \citenamefont {Gurarie},\ and\ \citenamefont {Rey}}]{Feig}%
  \BibitemOpen
  \bibfield  {author} {\bibinfo {author} {\bibfnamefont {M.}~\bibnamefont
  {Foss-Feig}}, \bibinfo {author} {\bibfnamefont {M.}~\bibnamefont {Hermele}},
  \bibinfo {author} {\bibfnamefont {V.}~\bibnamefont {Gurarie}}, \ and\
  \bibinfo {author} {\bibfnamefont {A.-M.}\ \bibnamefont {Rey}},\ }\href
  {\doibase 10.1103/PhysRevA.82.053624} {\bibfield  {journal} {\bibinfo
  {journal} {Phys. Rev. A}\ }\textbf {\bibinfo {volume} {82}},\ \bibinfo
  {pages} {053624} (\bibinfo {year} {2010}{\natexlab{a}})}\BibitemShut
  {NoStop}%
\bibitem [{\citenamefont {Jakobi}\ \emph {et~al.}(2013)\citenamefont {Jakobi},
  \citenamefont {Bl\"umer},\ and\ \citenamefont {van Dongen}}]{Jakobi}%
  \BibitemOpen
  \bibfield  {author} {\bibinfo {author} {\bibfnamefont {E.}~\bibnamefont
  {Jakobi}}, \bibinfo {author} {\bibfnamefont {N.}~\bibnamefont {Bl\"umer}}, \
  and\ \bibinfo {author} {\bibfnamefont {P.}~\bibnamefont {van Dongen}},\
  }\href {\doibase 10.1103/PhysRevB.87.205135} {\bibfield  {journal} {\bibinfo
  {journal} {Phys. Rev. B}\ }\textbf {\bibinfo {volume} {87}},\ \bibinfo
  {pages} {205135} (\bibinfo {year} {2013})}\BibitemShut {NoStop}%
\bibitem [{\citenamefont {Kugel'}\ and\ \citenamefont
  {Khomskii}(1982)}]{Kugel}%
  \BibitemOpen
  \bibfield  {author} {\bibinfo {author} {\bibfnamefont {K.~I.}\ \bibnamefont
  {Kugel'}}\ and\ \bibinfo {author} {\bibfnamefont {D.~I.}\ \bibnamefont
  {Khomskii}},\ }\href {http://stacks.iop.org/0038-5670/25/i=4/a=R03}
  {\bibfield  {journal} {\bibinfo  {journal} {Soviet Physics Uspekhi}\ }\textbf
  {\bibinfo {volume} {25}},\ \bibinfo {pages} {231} (\bibinfo {year}
  {1982})}\BibitemShut {NoStop}%
\bibitem [{\citenamefont {Foss-Feig}\ \emph
  {et~al.}(2010{\natexlab{b}})\citenamefont {Foss-Feig}, \citenamefont
  {Hermele},\ and\ \citenamefont {Rey}}]{Rey2009}%
  \BibitemOpen
  \bibfield  {author} {\bibinfo {author} {\bibfnamefont {M.}~\bibnamefont
  {Foss-Feig}}, \bibinfo {author} {\bibfnamefont {M.}~\bibnamefont {Hermele}},
  \ and\ \bibinfo {author} {\bibfnamefont {A.-M.}\ \bibnamefont {Rey}},\ }\href
  {\doibase 10.1103/PhysRevA.81.051603} {\bibfield  {journal} {\bibinfo
  {journal} {Phys. Rev. A}\ }\textbf {\bibinfo {volume} {81}},\ \bibinfo
  {pages} {051603(R)} (\bibinfo {year} {2010}{\natexlab{b}})}\BibitemShut
  {NoStop}%
\bibitem [{\citenamefont {Sotnikov}\ and\ \citenamefont
  {Hofstetter}(2014)}]{Sotnikov}%
  \BibitemOpen
  \bibfield  {author} {\bibinfo {author} {\bibfnamefont {A.}~\bibnamefont
  {Sotnikov}}\ and\ \bibinfo {author} {\bibfnamefont {W.}~\bibnamefont
  {Hofstetter}},\ }\href {\doibase 10.1103/PhysRevA.89.063601} {\bibfield
  {journal} {\bibinfo  {journal} {Phys. Rev. A}\ }\textbf {\bibinfo {volume}
  {89}},\ \bibinfo {pages} {063601} (\bibinfo {year} {2014})}\BibitemShut
  {NoStop}%
\bibitem [{\citenamefont {Zener}(1951)}]{Zener}%
  \BibitemOpen
  \bibfield  {author} {\bibinfo {author} {\bibfnamefont {C.}~\bibnamefont
  {Zener}},\ }\href {\doibase 10.1103/PhysRev.82.403} {\bibfield  {journal}
  {\bibinfo  {journal} {Phys. Rev.}\ }\textbf {\bibinfo {volume} {82}},\
  \bibinfo {pages} {403} (\bibinfo {year} {1951})}\BibitemShut {NoStop}%
\bibitem [{\citenamefont {Stellmer}\ \emph {et~al.}(2011)\citenamefont
  {Stellmer}, \citenamefont {Grimm},\ and\ \citenamefont
  {Schreck}}]{Ste2011PRA}%
  \BibitemOpen
  \bibfield  {author} {\bibinfo {author} {\bibfnamefont {S.}~\bibnamefont
  {Stellmer}}, \bibinfo {author} {\bibfnamefont {R.}~\bibnamefont {Grimm}}, \
  and\ \bibinfo {author} {\bibfnamefont {F.}~\bibnamefont {Schreck}},\ }\href
  {\doibase 10.1103/PhysRevA.84.043611} {\bibfield  {journal} {\bibinfo
  {journal} {Phys. Rev. A}\ }\textbf {\bibinfo {volume} {84}},\ \bibinfo
  {pages} {043611} (\bibinfo {year} {2011})}\BibitemShut {NoStop}%
\bibitem [{Note1()}]{Note1}%
  \BibitemOpen
  \bibinfo {note} {Private communication with S. F{\"o}lling, F. Scazza, D. R.
  Fernandes and L. Riegger, LMU, Munich.}\BibitemShut {Stop}%
\bibitem [{\citenamefont {Held}\ and\ \citenamefont {Vollhardt}(1998)}]{Held}%
  \BibitemOpen
  \bibfield  {author} {\bibinfo {author} {\bibfnamefont {K.}~\bibnamefont
  {Held}}\ and\ \bibinfo {author} {\bibfnamefont {D.}~\bibnamefont
  {Vollhardt}},\ }\href@noop {} {\bibfield  {journal} {\bibinfo  {journal}
  {Eur. Phys. J. B}\ }\textbf {\bibinfo {volume} {5}},\ \bibinfo {pages} {473}
  (\bibinfo {year} {1998})}\BibitemShut {NoStop}%
\bibitem [{\citenamefont {Peters}\ and\ \citenamefont
  {Pruschke}(2010)}]{Peters}%
  \BibitemOpen
  \bibfield  {author} {\bibinfo {author} {\bibfnamefont {R.}~\bibnamefont
  {Peters}}\ and\ \bibinfo {author} {\bibfnamefont {T.}~\bibnamefont
  {Pruschke}},\ }\href {\doibase 10.1103/PhysRevB.81.035112} {\bibfield
  {journal} {\bibinfo  {journal} {Phys. Rev. B}\ }\textbf {\bibinfo {volume}
  {81}},\ \bibinfo {pages} {035112} (\bibinfo {year} {2010})}\BibitemShut
  {NoStop}%
\bibitem [{\citenamefont {Kune\v{s}}(2015)}]{Kunes}%
  \BibitemOpen
  \bibfield  {author} {\bibinfo {author} {\bibfnamefont {J.}~\bibnamefont
  {Kune\v{s}}},\ }\href {http://stacks.iop.org/0953-8984/27/i=33/a=333201}
  {\bibfield  {journal} {\bibinfo  {journal} {J. Phys.: Condensed Matter}\
  }\textbf {\bibinfo {volume} {27}},\ \bibinfo {pages} {333201} (\bibinfo
  {year} {2015})}\BibitemShut {NoStop}%
\bibitem [{\citenamefont {Georges}\ \emph {et~al.}(1996)\citenamefont
  {Georges}, \citenamefont {Kotliar}, \citenamefont {Krauth},\ and\
  \citenamefont {Rozenberg}}]{Georges}%
  \BibitemOpen
  \bibfield  {author} {\bibinfo {author} {\bibfnamefont {A.}~\bibnamefont
  {Georges}}, \bibinfo {author} {\bibfnamefont {G.}~\bibnamefont {Kotliar}},
  \bibinfo {author} {\bibfnamefont {W.}~\bibnamefont {Krauth}}, \ and\ \bibinfo
  {author} {\bibfnamefont {M.~J.}\ \bibnamefont {Rozenberg}},\ }\href {\doibase
  10.1103/RevModPhys.68.13} {\bibfield  {journal} {\bibinfo  {journal} {Rev.
  Mod. Phys.}\ }\textbf {\bibinfo {volume} {68}},\ \bibinfo {pages} {13}
  (\bibinfo {year} {1996})}\BibitemShut {NoStop}%
\bibitem [{\citenamefont {Vollhardt}(2010)}]{Vollhardt}%
  \BibitemOpen
  \bibfield  {author} {\bibinfo {author} {\bibfnamefont {D.}~\bibnamefont
  {Vollhardt}},\ }\href@noop {} {\bibfield  {journal} {\bibinfo  {journal} {AIP
  Conference Proceedings}\ }\textbf {\bibinfo {volume} {1297}},\ \bibinfo
  {pages} {339} (\bibinfo {year} {2010})}\BibitemShut {NoStop}%
\bibitem [{\citenamefont {Isidori}(2007)}]{Isidori}%
  \BibitemOpen
  \bibfield  {author} {\bibinfo {author} {\bibfnamefont {A.}~\bibnamefont
  {Isidori}},\ }\emph {\bibinfo {title} {Superconductivity in Strongly
  Correlated Electron Systems: Analytical Approaches Beyond Dynamical Mean
  Field Theory}},\ \href@noop {} {Ph.D. thesis},\ \bibinfo  {school} {Sapienza
  University of Rome} (\bibinfo {year} {2007})\BibitemShut {NoStop}%
\bibitem [{\citenamefont {Anderson}(1961)}]{Anderson}%
  \BibitemOpen
  \bibfield  {author} {\bibinfo {author} {\bibfnamefont {P.}~\bibnamefont
  {Anderson}},\ }\href {\doibase 10.1103/PhysRev.124.41} {\bibfield  {journal}
  {\bibinfo  {journal} {Phys. Rev.}\ }\textbf {\bibinfo {volume} {124}},\
  \bibinfo {pages} {41} (\bibinfo {year} {1961})}\BibitemShut {NoStop}%
\bibitem [{\citenamefont {Caffarel}\ and\ \citenamefont
  {Krauth}(1994)}]{Caffarel}%
  \BibitemOpen
  \bibfield  {author} {\bibinfo {author} {\bibfnamefont {M.}~\bibnamefont
  {Caffarel}}\ and\ \bibinfo {author} {\bibfnamefont {W.}~\bibnamefont
  {Krauth}},\ }\href {\doibase 10.1103/PhysRevLett.72.1545} {\bibfield
  {journal} {\bibinfo  {journal} {Phys. Rev. Lett.}\ }\textbf {\bibinfo
  {volume} {72}},\ \bibinfo {pages} {1545} (\bibinfo {year}
  {1994})}\BibitemShut {NoStop}%
\bibitem [{\citenamefont {Sotnikov}(2015)}]{Sotnikov2}%
  \BibitemOpen
  \bibfield  {author} {\bibinfo {author} {\bibfnamefont {A.}~\bibnamefont
  {Sotnikov}},\ }\href {\doibase 10.1103/PhysRevA.92.023633} {\bibfield
  {journal} {\bibinfo  {journal} {Phys. Rev. A}\ }\textbf {\bibinfo {volume}
  {92}},\ \bibinfo {pages} {023633} (\bibinfo {year} {2015})}\BibitemShut
  {NoStop}%
\bibitem [{\citenamefont {Gull}\ \emph {et~al.}(2011)\citenamefont {Gull},
  \citenamefont {Millis}, \citenamefont {Lichtenstein}, \citenamefont
  {Rubtsov}, \citenamefont {Troyer},\ and\ \citenamefont {Werner}}]{Gull}%
  \BibitemOpen
  \bibfield  {author} {\bibinfo {author} {\bibfnamefont {E.}~\bibnamefont
  {Gull}}, \bibinfo {author} {\bibfnamefont {A.~J.}\ \bibnamefont {Millis}},
  \bibinfo {author} {\bibfnamefont {A.~I.}\ \bibnamefont {Lichtenstein}},
  \bibinfo {author} {\bibfnamefont {A.~N.}\ \bibnamefont {Rubtsov}}, \bibinfo
  {author} {\bibfnamefont {M.}~\bibnamefont {Troyer}}, \ and\ \bibinfo {author}
  {\bibfnamefont {P.}~\bibnamefont {Werner}},\ }\href {\doibase
  10.1103/RevModPhys.83.349} {\bibfield  {journal} {\bibinfo  {journal} {Rev.
  Mod. Phys.}\ }\textbf {\bibinfo {volume} {83}},\ \bibinfo {pages} {349}
  (\bibinfo {year} {2011})}\BibitemShut {NoStop}%
\bibitem [{\citenamefont {Buchhold}(2012)}]{Buchhold}%
  \BibitemOpen
  \bibfield  {author} {\bibinfo {author} {\bibfnamefont {M.}~\bibnamefont
  {Buchhold}},\ }\href@noop {} {\enquote {\bibinfo {title} {Topological phases
  of interacting fermions in optical lattices with artificial gauge fields},}\
  } (\bibinfo {year} {2012})\BibitemShut {NoStop}%
\bibitem [{\citenamefont {Helmes}\ \emph {et~al.}(2008)\citenamefont {Helmes},
  \citenamefont {Costi},\ and\ \citenamefont {Rosch}}]{Helmes}%
  \BibitemOpen
  \bibfield  {author} {\bibinfo {author} {\bibfnamefont {R.~W.}\ \bibnamefont
  {Helmes}}, \bibinfo {author} {\bibfnamefont {T.~A.}\ \bibnamefont {Costi}}, \
  and\ \bibinfo {author} {\bibfnamefont {A.}~\bibnamefont {Rosch}},\ }\href
  {\doibase 10.1103/PhysRevLett.100.056403} {\bibfield  {journal} {\bibinfo
  {journal} {Phys. Rev. Lett.}\ }\textbf {\bibinfo {volume} {100}},\ \bibinfo
  {pages} {056403} (\bibinfo {year} {2008})}\BibitemShut {NoStop}%
\bibitem [{\citenamefont {Snoek}\ \emph {et~al.}(2008)\citenamefont {Snoek},
  \citenamefont {Titvinidze}, \citenamefont {T\"oke}, \citenamefont {Byczuk},\
  and\ \citenamefont {Hofstetter}}]{Snoek}%
  \BibitemOpen
  \bibfield  {author} {\bibinfo {author} {\bibfnamefont {M.}~\bibnamefont
  {Snoek}}, \bibinfo {author} {\bibfnamefont {I.}~\bibnamefont {Titvinidze}},
  \bibinfo {author} {\bibfnamefont {C.}~\bibnamefont {T\"oke}}, \bibinfo
  {author} {\bibfnamefont {K.}~\bibnamefont {Byczuk}}, \ and\ \bibinfo {author}
  {\bibfnamefont {W.}~\bibnamefont {Hofstetter}},\ }\href
  {http://stacks.iop.org/1367-2630/10/i=9/a=093008} {\bibfield  {journal}
  {\bibinfo  {journal} {New J. Phys.}\ }\textbf {\bibinfo {volume} {10}},\
  \bibinfo {pages} {093008} (\bibinfo {year} {2008})}\BibitemShut {NoStop}%
\bibitem [{\citenamefont {Kubo}\ \emph {et~al.}(1999)\citenamefont {Kubo},
  \citenamefont {Edwards}, \citenamefont {Green}, \citenamefont {Momoi},\ and\
  \citenamefont {Sakamoto}}]{Kubo}%
  \BibitemOpen
  \bibfield  {author} {\bibinfo {author} {\bibfnamefont {K.}~\bibnamefont
  {Kubo}}, \bibinfo {author} {\bibfnamefont {D.}~\bibnamefont {Edwards}},
  \bibinfo {author} {\bibfnamefont {A.}~\bibnamefont {Green}}, \bibinfo
  {author} {\bibfnamefont {T.}~\bibnamefont {Momoi}}, \ and\ \bibinfo {author}
  {\bibfnamefont {H.}~\bibnamefont {Sakamoto}},\ }\href@noop {} {\emph
  {\bibinfo {title} {Physics of Manganites}}}\ (\bibinfo  {publisher} {Plenum
  Publishers},\ \bibinfo {year} {1999})\BibitemShut {NoStop}%
\bibitem [{\citenamefont {Werner}\ \emph {et~al.}(2005)\citenamefont {Werner},
  \citenamefont {Parcollet}, \citenamefont {Georges},\ and\ \citenamefont
  {Hassan}}]{Werner}%
  \BibitemOpen
  \bibfield  {author} {\bibinfo {author} {\bibfnamefont {F.}~\bibnamefont
  {Werner}}, \bibinfo {author} {\bibfnamefont {O.}~\bibnamefont {Parcollet}},
  \bibinfo {author} {\bibfnamefont {A.}~\bibnamefont {Georges}}, \ and\
  \bibinfo {author} {\bibfnamefont {S.~R.}\ \bibnamefont {Hassan}},\ }\href
  {\doibase 10.1103/PhysRevLett.95.056401} {\bibfield  {journal} {\bibinfo
  {journal} {Phys. Rev. Lett.}\ }\textbf {\bibinfo {volume} {95}},\ \bibinfo
  {pages} {056401} (\bibinfo {year} {2005})}\BibitemShut {NoStop}%
\bibitem [{\citenamefont {Golubeva}\ \emph {et~al.}(2015)\citenamefont
  {Golubeva}, \citenamefont {Sotnikov},\ and\ \citenamefont
  {Hofstetter}}]{Golubeva}%
  \BibitemOpen
  \bibfield  {author} {\bibinfo {author} {\bibfnamefont {A.}~\bibnamefont
  {Golubeva}}, \bibinfo {author} {\bibfnamefont {A.}~\bibnamefont {Sotnikov}},
  \ and\ \bibinfo {author} {\bibfnamefont {W.}~\bibnamefont {Hofstetter}},\
  }\href {\doibase 10.1103/PhysRevA.92.043623} {\bibfield  {journal} {\bibinfo
  {journal} {Phys. Rev. A}\ }\textbf {\bibinfo {volume} {92}},\ \bibinfo
  {pages} {043623} (\bibinfo {year} {2015})}\BibitemShut {NoStop}%
\bibitem [{\citenamefont {Inaba}\ and\ \citenamefont {Suga}(2013)}]{Inaba}%
  \BibitemOpen
  \bibfield  {author} {\bibinfo {author} {\bibfnamefont {K.}~\bibnamefont
  {Inaba}}\ and\ \bibinfo {author} {\bibfnamefont {S.}~\bibnamefont {Suga}},\
  }\href {\doibase 10.1142/S0217984913300081} {\bibfield  {journal} {\bibinfo
  {journal} {Mod. Phys. Lett.}\ }\textbf {\bibinfo {volume} {B27}},\ \bibinfo
  {pages} {1330008} (\bibinfo {year} {2013})}\BibitemShut {NoStop}%
\bibitem [{\citenamefont {Zhang}\ \emph {et~al.}(2016)\citenamefont {Zhang},
  \citenamefont {Zhang}, \citenamefont {Cheng}, \citenamefont {Chen},
  \citenamefont {Zhang},\ and\ \citenamefont {Zhai}}]{Zhang}%
  \BibitemOpen
  \bibfield  {author} {\bibinfo {author} {\bibfnamefont {R.}~\bibnamefont
  {Zhang}}, \bibinfo {author} {\bibfnamefont {D.}~\bibnamefont {Zhang}},
  \bibinfo {author} {\bibfnamefont {Y.}~\bibnamefont {Cheng}}, \bibinfo
  {author} {\bibfnamefont {W.}~\bibnamefont {Chen}}, \bibinfo {author}
  {\bibfnamefont {P.}~\bibnamefont {Zhang}}, \ and\ \bibinfo {author}
  {\bibfnamefont {H.}~\bibnamefont {Zhai}},\ }\href {\doibase
  10.1103/PhysRevA.93.043601} {\bibfield  {journal} {\bibinfo  {journal} {Phys.
  Rev. A}\ }\textbf {\bibinfo {volume} {93}},\ \bibinfo {pages} {043601}
  (\bibinfo {year} {2016})}\BibitemShut {NoStop}%
\bibitem [{\citenamefont {Cheng}\ \emph {et~al.}(2016)\citenamefont {Cheng},
  \citenamefont {Zhang},\ and\ \citenamefont {Zhang}}]{Zhang-2}%
  \BibitemOpen
  \bibfield  {author} {\bibinfo {author} {\bibfnamefont {Y.}~\bibnamefont
  {Cheng}}, \bibinfo {author} {\bibfnamefont {R.}~\bibnamefont {Zhang}}, \ and\
  \bibinfo {author} {\bibfnamefont {P.}~\bibnamefont {Zhang}},\ }\href@noop {}
  {\bibfield  {journal} {\bibinfo  {journal} {Phys. Rev. A}\ }\textbf {\bibinfo
  {volume} {93}},\ \bibinfo {pages} {042708} (\bibinfo {year}
  {2016})}\BibitemShut {NoStop}%
\bibitem [{\citenamefont {Scazza}(2015)}]{ScazzaPhD}%
  \BibitemOpen
  \bibfield  {author} {\bibinfo {author} {\bibfnamefont {F.}~\bibnamefont
  {Scazza}},\ }\emph {\bibinfo {title} {Probing SU(N)-symmetric orbital
  interactions with ytterbium Fermi gases in optical lattices}},\ \href@noop {}
  {Ph.D. thesis},\ \bibinfo  {school} {Ludwig-Maximilians-Universit\"at
  M\"unchen} (\bibinfo {year} {2015})\BibitemShut {NoStop}%
\end{thebibliography}%

\end{document}